\documentclass[%
 reprint,
 amsmath,amssymb,
 aps,
]{revtex4-2}
\usepackage{graphicx}
\usepackage{dcolumn}
\usepackage{bm}
\usepackage{color}

\newcommand{\xzy}[1]{\textcolor{black}{ #1}}

\begin{document}


\title{Extracting Astrophysical Information of Highly-Eccentric Binaries in the Millihertz Gravitational Wave Band
}

\author{Zeyuan Xuan}
\email{zeyuan.xuan@physics.ucla.edu}
\affiliation{ Department of Physics and Astronomy, UCLA, Los Angeles, CA 90095}
\affiliation{Mani L. Bhaumik Institute for Theoretical Physics, Department of Physics and Astronomy, UCLA, Los Angeles, CA 90095, USA}

\author{Smadar Naoz}
\affiliation{ Department of Physics and Astronomy, UCLA, Los Angeles, CA 90095}
\affiliation{Mani L. Bhaumik Institute for Theoretical Physics, Department of Physics and Astronomy, UCLA, Los Angeles, CA 90095, USA}

\author{Alvin K. Y. Li}
\affiliation{LIGO Laboratory, California Institute of Technology, Pasadena, CA 91125, USA}
\affiliation{Department of Physics, The Chinese University of Hong Kong, Hong Kong}
\affiliation{RESCEU, The University of Tokyo, Tokyo, 113-0033, Japan}

\author{Bence Kocsis}
\affiliation{Rudolf Peierls Centre for Theoretical Physics, Parks Road, Oxford OX1 3PU, UK}

\author{Erik Petigura}
\affiliation{Department of Physics and Astronomy, UCLA, Los Angeles, CA 90095}

\author{Alan M. Knee}
\affiliation{Department of Physics and Astronomy, University of British Columbia, Vancouver, BC V6T 1Z1, Canada}
\author{Jess McIver}
\affiliation{Department of Physics and Astronomy, University of British Columbia, Vancouver, BC V6T 1Z1, Canada}
\author{Kyle Kremer}
\affiliation{Department of Astronomy and Astrophysics, University of California, San Diego, 9500 Gilman Drive, La Jolla, CA, 92093, USA}

\author{Will M. Farr}
\affiliation{Center for Computational Astrophysics, Flatiron Institute, 162 Fifth Avenue, New York, NY 10010, USA}
\affiliation{Department of Physics and Astronomy, Stony Brook University, Stony Brook NY 11794, USA}

\date{\today}

\begin{abstract}
Wide, highly eccentric ($e>0.9$) compact binaries can naturally arise as progenitors of gravitational wave (GW) mergers. These systems are expected to have a significant population in the mHz band \xzy{(e.g., $\sim 3-45$ detectable stellar-mass binary black holes with $e>0.9$ in the Milky Way)}, with their GW signals characterized by ``repeated bursts" emitted upon each pericenter passage. 
In this study, we show that the detection of mHz GW signals from highly eccentric stellar mass binaries in the local universe can strongly constrain their orbital parameters. Specifically, it can achieve a relative measurement error of $\sim 10^{-6}$ for orbital frequency and $\sim 1\%$ for eccentricity (as $1-e$) in most of the detectable cases.
On the other hand, the binary's mass ratio, distance, and intrinsic orbital orientation may be less precisely determined due to degeneracies in the GW waveform.  We also perform mock LISA data analysis to evaluate the realistic detectability of highly eccentric compact binaries. Our results show that highly eccentric systems could be efficiently identified when multiple GW sources and stationary Gaussian instrumental noise are present in the detector output. 
This work highlights the potential of extracting the signal of ``bursting'' LISA sources to provide valuable insights into their orbital evolution, surrounding environment, and formation channels.

\end{abstract}

\maketitle

\section{Introduction}
\label{section:intro}

Many compact object binaries are expected to have non-negligible eccentricity in the millihertz gravitational wave (GW) band. In particular, GW sources formed via dynamical channels can naturally undergo a progenitor stage before the final merger, during which the environment perturbs the binary and excites the orbital eccentricity (e.g., $e\gtrsim 0.9$); therefore, even if the GW source is initially characterized by a large semi-major axis (e.g., $a\gtrsim 0.1 \,\rm au$), its pericenter distance, $r_p=a(1-e)$, can become sufficiently small to induce strong millihertz GW emission \citep[][]{Loutrel_2017,Loutrel+20,Xuan+23b}, leading to orbital energy loss and eventually resulting in a GW merger. For example, in dense star clusters, external perturbations like GW capture, binary-single, and binary-binary scattering \citep[e.g.,][]{O'Leary+09,Thompson+11,Aarseth+12,Kocsis_2012,breivik16,Gondan_2018a,Orazio+18,Zevin_2019,Samsing+19,Stephan+16,Hoang+20,Martinez+20,Antonini+19,Kremer_2020,wintergranic2023binary,Gondan_Kocsis2021,Zhang+21,Fragione+22clustermerger,Samsing+2022,Rowan+23agn,purohit2024binary} can drive a significant fraction of wide compact object binaries into highly eccentric orbits. In a hierarchical triple system, where a tight binary orbits a third body on a much wider ``outer orbit", the inner binary can undergo eccentricity oscillations due to the eccentric Kozai-Lidov (EKL) mechanism \citep{Kozai1962,Lidov1962,Ligongjie+15ekl,Naoz16,Hoang+18}, potentially leading to an eccentric GW merger. Furthermore, eccentric compact binaries could form in stellar disks or active galactic nucleus accretion disks  \citep{tagawa+20agn,Tagawa+2021,Samsing+2022,Munoz+22,Gautham+23,Arca+23,Yan+23highspeedsource,peng+23EMRIAGN,Whitehead+24agngas,Zhu+24AGN}. In the galactic field, fly-by interactions and galactic tides may also produce eccentric GW sources \citep[e.g.,][]{Michaely+19,Michaely+20,Michaely+22,Jakob24flyby}. \xzy{These eccentric binaries can significantly contribute to the number of sources in mHz GW detection (e.g., $\sim 3-45$ detectable stellar-mass binary black holes with $e>0.9$ in the Milky Way \citep{Xuan+23b}), and yield non-negligible merger rate of compact objects in the Local Group \citep[][]{wen03,Hoang+18,Hamers+18,Stephan+19,Zevin_2019,Bub+20,Deme+20,Wang+21,Zevin_2021} over the expected observation time of the future Laser
Interferometer Space Antenna (LISA) mission \citep{2017arXiv170200786A}.}

So far, many studies have focused on measuring the residual eccentricity of GW mergers detected by LIGO, Virgo, and KAGRA (LVK), which could significantly enhance our understanding of the formation mechanisms of compact binaries \citep[][]{east13, samsing14, Coughlin_2015, Gondan_2018a, Gondan_2018b, moore19,2021ApJ...913L...7A, 2021arXiv211103634T,Zevin_2021,Lower18, Romero_Shaw_2019,Knee2022ecc}. However, there is still a lack of clear 
observational evidence for the eccentricity of GW sources \citep[see, e.g.,][]{Abbott_2019ecc,Lenon_2020, Romero-Shaw_2020,gayathri2022eccentricity,Samsing+2022}, mostly because GW radiation tends to circularize the orbit, rendering eccentricity negligible within the sensitive frequency band of current detectors \citep{Peters64,Hiner08}. On the other hand, LISA \citep{2017arXiv170200786A} will observe sources in a lower frequency band ($10^{-4}-10^{-1}~\rm Hz$), allowing us to probe the earlier evolutionary stages of these eccentric GW mergers \citep[see, e.g., ][]{barack04,Mikoczi+12,robson18, chen19, Hoang+19, Fang19, tamanini19, Torres-Orjuela21, amaro+22,Xuan+21, Xuan23acc}. Therefore, it is crucial to evaluate the detectability and parameter measurement accuracy of eccentric GW sources, particularly for mHz detections by LISA.

This work focuses on the astrophysical inference of highly eccentric, stellar-mass compact object binaries. Specifically, the GW signal emitted by wide, highly eccentric compact binaries has a unique signature in the mHz band. For example, when a binary's eccentricity is small, its GW signal can be approximated by a near-monochromatic, sinusoidal wave \citep[see, e.g., ][]{Cutler+94}. However, as the eccentricity increases, the GW emission becomes stronger upon each pericenter passage, transforming the signal into a burst-like waveform \citep[e.g.,][]{Kocsis_2012,Xuan+23b,Xuan24bkg}. These bursting GW signals, characterized by transient pulses in the time domain, generate a frequency power spectrum with numerous harmonics (on the order of $\sim 10^3-10^6$) that contribute significantly to the total signal-to-noise ratio \citep[see, e.g., ][]{Kocsis_2012,Xuan+23b}. As a result, the energy of the GW signal is spread across a wide frequency range, making it necessary to employ methods that can resolve broadband transient signals, such as wavelet decomposition \citep{Klimenko_2004,Chatterji2004Qtransform,Chatterji2006searchburst,bassetti2005development,Tai_2014}.
Furthermore, multiple bursting sources may simultaneously contribute to the signal \citep{Xuan+23b}, highlighting the need to identify and disentangle individual sources, as demonstrated below.  %
The collective GW signal from multiple highly eccentric binaries could also form a stochastic GW background, where numerous GW harmonics from different sources overlap \citep{Xuan24bkg}, and interfere with our understanding of other LISA sources. 

It has been suggested that eccentric GW signals can be detected in the LISA band, leading to extracting information about the astrophysical sources \citep[see, e.g., ][]{Seto+01,Glampedakis_2005,Hopman2006,Rubbo06,Amaro_Seoane_2007,Barack_2009,Thompson+11,Mikoczi+12,Berry_2013,Chen_2018,Fang_2019,Hoang+19,Wang+21,Randall_2022,Fan22,oliver2023gravitational,Naoz+22,Naoz+23,Zhang_2021,Xuan23acc,Shariat+23WDgaia,nishizawa16a,nishizawa16b}. Furthermore, many efforts have been made to analyze the mHz bursting GW signals using various waveform templates \citep{Loutrel_2017,Moore2018,Han+20EMRIburst,Loutrel+20,Romero23, knee2024detectinggravitationalwaveburstsblack,Zhangzhongfu+24flybyburst}. However, it remains unclear whether we can successfully distinguish highly eccentric ($e>0.9$) sources in mHz data analysis and what level of parameter estimation accuracy can be realistically achieved for such systems.
Below, we quantify the parameter measurement accuracy using a Fisher matrix analysis and, for the first time, demonstrate the application of matched filtering for identifying highly eccentric stellar-mass binaries in the LISA band, in the presence of multiple GW sources and instrumental detector noise.
In this work, we analyze the bursting compact binaries in the Milky Way as a representative example. However, the results of Fisher matrix analysis and matched filtering can be generalized, with a rescaling of distance and signal-to-noise ratio, to bursting sources in the local universe with negligible redshift.

Note that the Fisher matrix analysis \citep{Coe+09, Cutler+94} has been widely used in the literature to evaluate the parameter measurement accuracy for LISA, because of its simplicity and computational efficiency. However, this method can sometimes provide inaccurate results, particularly for waveforms with low expected signal-to-noise ratios (SNR) or for signals that depend weakly on certain parameters \citep[see, e.g.,][]{Vallisneri_2008, Toubiana_2020}. Therefore, the parameter measurement accuracy presented in our work should be interpreted as an initial estimation for this new class of sources (i.e., wide, highly eccentric binaries with slow orbital evolution). A Bayesian analysis is beyond the scope of this paper.

The paper is organized as follows: In Section~\ref{sec:astro}, for completeness,
we reviewed the astrophysical properties of highly eccentric GW sources \citep{Xuan+23b, Xuan24bkg}. Next, we discussed the waveform model used in this work and 
described the parameterization of the waveform template (see Section~\ref{sec:waveform_model}). In Section~\ref{sec:fisherana}, we estimated LISA's parameter measurement accuracy for highly eccentric GW sources using the Fisher matrix analysis. Furthermore, we carried out mock LISA data analysis to 
discuss the identification of bursting sources under realistic assumptions. In particular, we introduce the matched filtering method in Section~\ref{sec:MFmethod}, which is commonly adopted to analyze GW data, then highlight the difference between the data analysis of bursting and other quasi-circular GW sources (see Section~\ref{sunsec:differenceofburstfitting}). The 
simulation of mock LISA data is discussed in Section~\ref{subsec: population}, and the mock analysis results are shown in Section~\ref{subsec:mockobsresult}. In Section~\ref{sec:discussion}, we summarized the paper and discussed the astrophysical implications. Throughout the paper, unless otherwise specified, we set $G=c=1$.

\section{Astrophysical properties and waveform modeling}
\subsection{Astrophysical Properties}\label{sec:astro}
As discussed in Section~\ref{section:intro}, wide, highly eccentric compact binaries often serve as progenitors of GW mergers, particularly in dynamical channels where environmental perturbations cause close encounters between two compact objects. Furthermore, as shown in our previous works \citep{Xuan+23b,Xuan24bkg}, these highly eccentric binaries could have promising prospects for detectability, extended lifetime, and a significant population in the local universe. Therefore, for completeness, we briefly summarize the relevant findings and discuss the signatures of highly eccentric waveforms in the mHz band in this section.

The GW emission from a highly eccentric binary is largely suppressed for most of the orbital period. However, during each pericenter passage, the distance between the two components of the binary will decrease significantly, 
producing a sudden burst of GW radiation. Thus, the GW signal from wide, highly eccentric sources is characterized by ``repeated bursts" \citep{kocsis06,Kocsis_2012,Tai_2014,Loutrel_2017,Gondan_2018b,fangx19,knee2024detectinggravitationalwaveburstsblack}, where the separation between two bursts is the orbital period. Furthermore, the duration of each GW burst is approximately the pericenter passage time, $T_p$ \citep[][]{O'Leary+09,Xuan+23b}, which can be estimated using the pericenter distance divided by the orbital velocity at pericenter:
\begin{equation}
   T_{p} \sim \frac{r_p}{v_p} \sim (1-e)^{3/2} T_{\rm orb} \ ,
    \label{eq:time0}
\end{equation}
where $v_p$ is the orbital velocity at pericenter, and $T_{\rm orb}=2\pi a^{3/2} M^{-1/2}$ is the period of a binary with a total mass $M$. Note that we omit an order unity factor of $(1+e)^{-1/2}$, following our previous work \citep{Xuan+23b}.

Furthermore, we can estimate the strain amplitude, $h_{\rm burst}$, and peak frequency, $f_{\rm burst}$, of a single GW pulse in the waveform of a highly eccentric compact binary using the following relations \citep{Xuan+23b}:
\begin{align}
    f_{\rm burst}&\sim 2f_{\rm orb}(1-e)^{-\frac{3}{2}} \nonumber\\
    &\sim 3.16 {\rm mHz}\, \left(\frac{M}{20\rm M_{\odot}}\right)^{\frac{1}{2}} \left(\frac{a}{1\rm au}\right)^{-\frac{3}{2}}\left(\frac{1-e}{0.002}\right)^{-\frac{3}{2}} \ ,
    \label{eq:fpeak}
\end{align}
and
\begin{align}
    h_{\rm burst} &\sim \sqrt{\frac{32}{5}}\frac{m_1m_2}{Ra(1-e)}\nonumber\\
    &\sim  7.65\times 10^{-21}\eta_s \left(\frac{M}{20\rm M_{\odot}}\right)^{\frac{5}{3}} \left(\frac{f_{\rm burst}}{3.16\rm mHz}\right)^{\frac{2}{3}}\left(\frac{R}{8\rm kpc}\right)^{-1}\ ,
        \label{eq:hburst}
\end{align}
in which $f_{\rm orb}=1/T_{\rm orb}$ is the orbital frequency of the bursting source, $R$ is the distance of the binary from the observer, $m_1,\, m_2$ are the mass of the binary's components, and $\eta_s=4 m_1m_2/(m_1+m_2)^{2}$ is unity for equal mass sources. We note that the peak GW frequency of eccentric sources is often estimated as $f_{\rm peak} = f_{\rm orb} (1+e)^{1/2} (1-e)^{-3/2}$ \citep{O'Leary+09}. A more detailed expression of the peak frequency can also be found in Refs.~\citep{wen03,Randall_2022}. For consistency with other definitions in our previous treatment \citep{Xuan+23b}, here we adopt $f_{\rm burst}\sim 2 f_{\rm orb} (1-e)^{-3/2}$, which differs by an order of unity.

As shown in Equations~(\ref{eq:fpeak}) and (\ref{eq:hburst}), even when the binary is considerably wide and the orbital frequency is well below the LISA band (e.g., $a \sim 1 \rm au$), an increase in the orbital eccentricity can result in mHz GW burst emission, with the strain amplitude strong enough to be detected by LISA. In particular, we can estimate the signal-to-noise ratio of these bursting sources using the following analytical equation \citep{Xuan+23b}\footnote{Note that the second line of Equation~(\ref{eq:SNR}) differs from Eq.11 in Ref.~\citep{Xuan+23b} by a factor of $\sqrt{2}$, which has been introduced to ensure consistency between the first and second lines. }:
\begin{align}
{\rm SNR} \sim
\left\{
\begin{array}{cc}
     \dfrac{h_{\rm burst}}{\sqrt{S_n\left(f_{\rm burst }\right)}} \sqrt{T_{\rm o b s}\left(1-e\right)^{3 / 2}} & (T_{\rm obs} \geq T_{\rm orb})\\[3ex]
     \dfrac{h_{\rm burst}}{\sqrt{f_{\rm burst} S_n(f_{\rm burst })/2}} & (T_{\rm obs} \leq T_{\rm orb})
\end{array}
\right.
\label{eq:SNR}
\end{align}
here $T_{\rm obs}$ is the observational time, and $S_{\mathrm{n}}(f)$ is the
spectral noise density of LISA evaluated at GW frequency $f$
\citep[see e.g.,][]{thorne_1987,Klein+16,2016PhRvD..93b4003K,Robson+19LISAnoise,} (Note that Equation~(\ref{eq:SNR}) is an approximation. For a more precise expression of the $\rm SNR$, see, for example, Appendix B of Ref.~\citep{Xuan+23b}.).

We highlight that, for a bursting source, the strong GW emission only happens for a short amount of time near the pericenter passage, while the binary will remain quiescent for the rest of the orbital period. Therefore, the overall energy emission and 
SNR will be suppressed by a factor of $\sim (1-e)^{3/2}$ compared to circular binaries with the same GW frequency (see, e.g., Equation~(\ref{eq:SNR})). However, such suppression in energy emission also results in a much slower orbital shrinkage and a significantly longer detectable time within the LISA band. In particular, we can estimate the lifetime of a bursting source, $\tau_{\rm burst}$, by considering the merger timescale of binaries with extreme eccentricity \citep{Peters64,Xuan+23b}:  
\begin{align}
    \tau_{\mathrm{burst}}\sim & \frac{3}{85\mu M^2} a^4\left(1-e^2\right)^{7 / 2} 
    \sim  1.17\times10^{6}{\rm yr}\times \nonumber\\&\,\eta_s^{-1}\left(\frac{M}{20\rm M_{\odot}}\right)^{-\frac{5}{3}} \left(\frac{f_{\rm burst}}{3.16\rm mHz}\right)^{-\frac{8}{3}}\left(\frac{1-e}{0.002}\right)^{-\frac{1}{2}} ,
    \label{eq:lifetime}
\end{align}
where $\mu=m_{1}m_{2}/(m_1+m_2)$. 

As shown in Equation~(\ref{eq:lifetime}), with the parameters of $a\sim 1$~au, $ e\sim0.998$, $m_1=m_2=10{\rm M_{\odot}}$, $R=8$~kpc, a stellar-mass bursting sources can remain detectable in the mHz band for $\sim10^6$~years \xzy{(note that this system has a signal-to-noise ratio $\rm SNR\sim$~50 for a 3~yr LISA observation; see Equation~(\ref{eq:SNR}))}. Such a lifetime is much longer than a stellar-mass circular binary's merger timescale, which is typically $\sim 10^3-10^4$~years in the mHz band \citep{Peters64}. Therefore, as a progenitor stage of GW mergers, the extended lifetime of highly eccentric binary indicates their large population in the local universe. For future observation, we expect LISA to detect many slowly evolving bursting sources as a natural consequence of compact binaries' dynamical formation. 

\begin{figure}[htbp]
    \centering
    \includegraphics[width=3.5in]{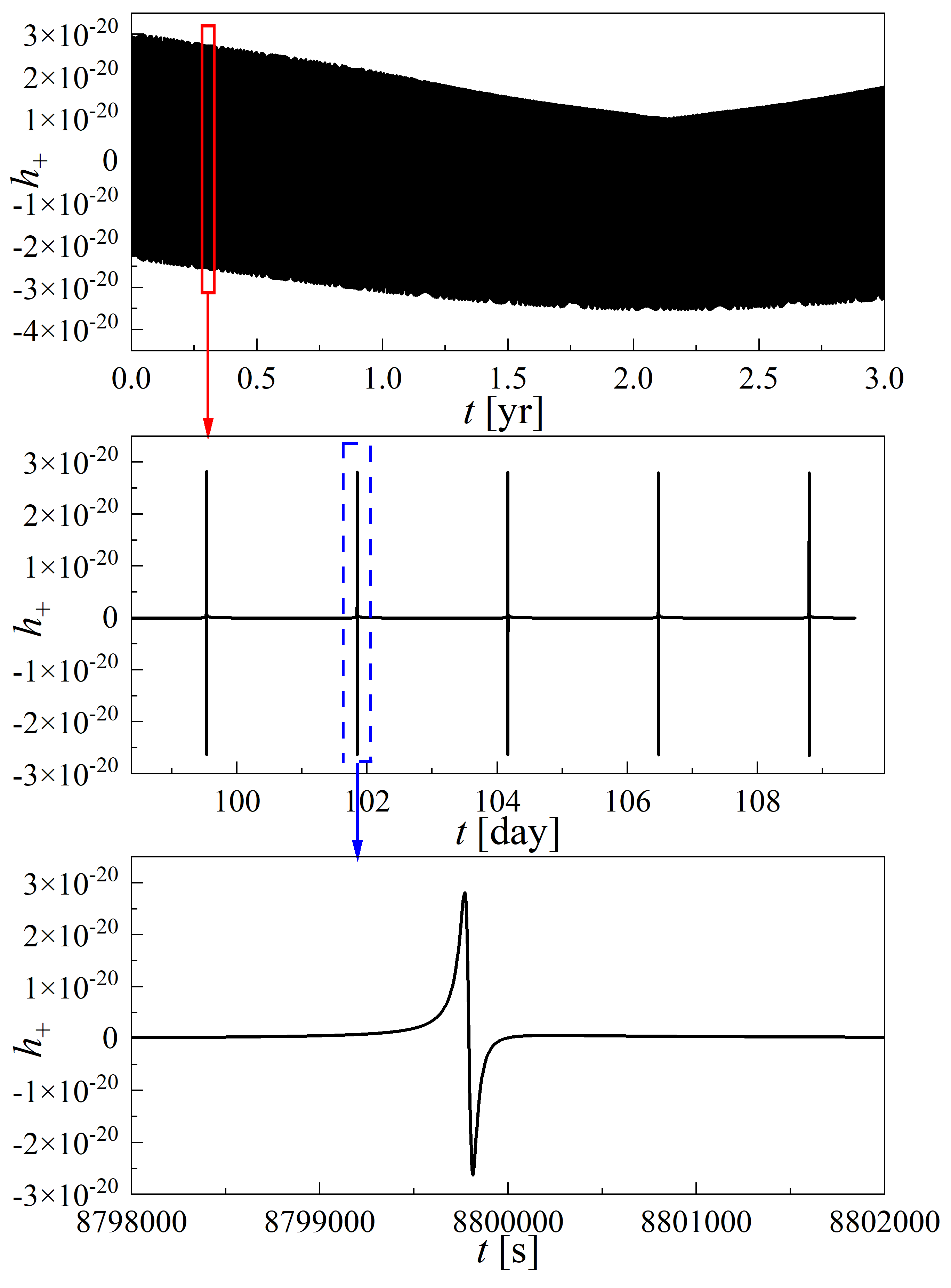} 
    \caption{{\bf Bursting GW waveform from a wide, highly eccentric BBH system.} Here we show the strain amplitude ($h_{+}$) of a BBH system with \xzy{$m_{1}=15$~M$_{\odot},$ $m_{2}=10$}~M$_{\odot},\,$$a= 0.100148 {\rm au},\,$$e=0.99,\,$$\Theta=\Phi=\pi/4$, placed at
    $R=8$~kpc and observed for $3$~years (see Section~\ref{sec:waveform_model} for detailed definitions of $\Theta,\Phi$). The time domain waveform is generated using the x-model as described in Section~\ref{sec:waveform_model}. {\it Upper Panel} shows the entire 3-year-long signal, which is made up of $\sim473$ GW bursts (clustering in the black-colored region). Furthermore, the amplitude modulation in the {\it Upper Panel} is caused by general relativistic precession of the binary's orbit (precession period $\sim 11.8$~yr). {\it Middle Panel} is a zoom-in of the signal for a $\sim 10$~days period, and {\it Bottom Panel} illustrates the details of waveform near a single GW burst. Note that in this example, we do not include the detector's response, otherwise, the LISA's annual motion around the sun will induce another long-term modulation on the amplitude and phase of the bursting signal. 
    }
    \label{fig:egwaveform}
\end{figure}

In Figure~\ref{fig:egwaveform}, we illustrate a representative example of bursting GW signals from highly eccentric, stellar-mass binary black hole (BBH) systems. Specifically, we choose the parameters of $m_{1}=15$~M$_{\odot},$ $m_{2}=10$~M$_{\odot},\,$$a= 0.100148 {\rm au},\,$$e=0.99,\,$$\Theta=\Phi=\pi/4$ (note that $\Theta,\Phi$ is the spherical polar
angles of the observer's direction as viewed in the source's comoving frame, see Figure~\ref{fig:coordinate} and Section~\ref{sec:waveform_model} for more details). This system is placed at $R=8$~kpc and observed for $3$~years, which stand for representative parameters of the bursting sources in the Milky Way \citep[see, e.g., the population analysis in ][]{kremer18,Xuan+23b,Xuan24bkg}. Furthermore, for this system, the $\rm SNR$ of a single GW burst is $\sim 62$ and the total $\rm SNR$ is $\sim 1353$ (see Equation~(\ref{eq:SNR})). However, in realistic observations, some detectable bursting sources may have the $\rm SNR$ of a single GW burst below the detection threshold (see, e.g., Ref.~\citep{Xuan+23b}). The time domain waveform is generated using the x-model as described in Section~\ref{sec:waveform_model}, and we choose to show the plus polarization, $h_{+}$, for demonstration purposes. 

We note that, the black-colored region in the {\it Upper Panel} of Figure~\ref{fig:egwaveform} represents a clustering of $\sim 473$ GW bursts, with the long-term amplitude modulation of their envelope caused by the general relativistic precession. Furthermore, due to the slow orbital evolution (see Equation~(\ref{eq:lifetime})), the orbital frequency, $f_{\rm orb}$, of the binary is approximately constant during the observation period.
Therefore, the waveform is characterized by repeated bursts with almost uniform separation (see {\it Middle Panel}). Additionally, as shown in {\it Bottom Panel}, a single GW burst roughly lasts hundreds of seconds, which indicates that the burst frequency lies in the mHz band. As shown in Figure~\ref{fig:egwaveform}, the ``bursting" waveform of highly eccentric binaries has a unique signature compared with other quasi-circular waveforms. Thus, it can provide us with valuable insight into the binary's dynamical evolution, as well as test the theory of gravity. In the following sections, we will discuss how to practically extract the astrophysical information of these sources from simulated LISA data.

\subsection{Waveform Model}\label{sec:waveform_model}

In this paper, we adopt the x-model \citep{Hinder+10} to generate the gravitational wave signal from eccentric binaries (assuming that the binaries are not perturbed during observation, and thus can be treated as isolated two-body systems). The x-model is a time-domain, post-Newtonian (pN)-based waveform family that captures all the critical features introduced by eccentricity in non-spinning binaries \citep{Huerta+14}. The binary orbit is parameterized using Keplerian parameterization to 3pN order, with the conservative evolution also given to 3pN order. In the x-model, the loss of energy and angular momentum is mapped to changes in the orbital eccentricity $e$ and the pN expansion parameter $x \equiv (\omega M)^{2/3}$, where $\omega$ is the mean Keplerian orbital frequency. These two parameters are evolved according to 2~pN equations. The x-model has been validated against numerical relativity for the case of equal mass BBHs with $e=0.1$, 21 cycles before the merger. It also agrees with well-studied template families in GW data analysis for the zero-eccentricity case \citep{Brown+10}.

In particular, we use Eqs. 9 - 10 in \citet{Hinder+10} to generate the time evolution of $x$ and $e$ (note that in the PN formalism, there are three different definitions of eccentricity parameters ($e_r,\,e_\phi,\,e_t$), and the eccentricity here stands for $e_t$, see Refs.~\citep{Damour1985,Damour1986,Blanchet_2002,Hinder+10,Knee2022ecc}), then evolve the orbital phase $\varphi$ and orbital radius $r$ using their Eqs. 5 - 7. The strain amplitude of two polarizations, $h_{+}$ and $h_{\times}$, are given (to leading Newtonian order) by:
\begin{equation}
\begin{aligned}
&\begin{aligned}
h_{+}= & -\frac{M \eta}{R}\left\{( \operatorname { c o s } ^ { 2 } \Theta + 1 ) \left[\cos 2 \varphi^{\prime}\left(-\dot{r}^2+r^2 \dot{\varphi}^2+\frac{M}{r}\right)\right.\right. \\
& \left.\left.+2 r \dot{r} \dot{\varphi} \sin 2 \varphi^{\prime}\right]+\left(-\dot{r}^2-r^2 \dot{\varphi}^2+\frac{M}{r}\right) \sin ^2 \Theta\right\} \ ,
\end{aligned}\\
&\begin{aligned}
h_{\times}= & -\frac{2 M \eta}{R} \cos \Theta\left\{\left(-\dot{r}^2+r^2 \dot{\varphi}^2+\frac{M}{r}\right) \sin 2 \varphi^{\prime}\right. \\
& \left.-2 r \cos 2 \varphi^{\prime} \dot{r} \dot{\varphi}\right\} \ , \label{eq:strains}
\end{aligned}
\end{aligned}
\end{equation}
where $\eta=m_1m_2/(m_1+m_2)^2$, $\varphi^{\prime}=\varphi(t)-\Phi$, $\Theta$ and $\Phi$ are the spherical polar
angles of the observer (i.e., the line of sight direction viewed in the source's frame), and a dot denotes the time derivative (i.e., $\dot{r}=dr/dt,\,\dot{\varphi}=d\varphi/dt$). 

Hereafter, for simplicity, we set $\varphi(t=0)=0$, $e(t=0)=e_0$, and start the waveform at the time when the binary is undergoing the pericenter passage ($\dot{r}=0$, this choice fixes the initial pericenter direction in the binary's comoving frame). Note that the semi-major axis $a$ is not well defined in the pN waveform. However, to better convey the astrophysical interpretation, we define $r(t=0)=a_0(1-e_0)$ and parameterize the initial condition using the approximated ``initial semi-major axis", $a_0$, in the Newtonian limit.

To better understand eccentric GW signals, we can also decompose the time-domain waveform of an eccentric binary into different harmonics (see, e.g., \citep{peters63,oleary09,Kocsis_2012,Kocsis2012b} for a detailed explanation),
\begin{equation}
h(t) \sim \sum_{n=1}^{\infty} h_{n} \exp \left(2 \pi i f_n t\right).\label{eq:eccdecompose}
\end{equation}
where each harmonic is a sinusoidal signal with amplitude $h_n$ and a frequency of $f_n=nf_{\rm orb}\, (n=1,\,2,\,3\,...)$. This property is useful when deriving the detector's response to eccentric signals, which we will discuss subsequently. \xzy{Furthermore, when GW sources evolve slowly (merger timescale $\tau\sim f_{\rm orb}/\dot{f}_{\rm orb}\gg T_{\rm obs}$), the orbital frequency $f_{\rm orb}$ almost keeps constant during observation. In this case, the harmonics in Equation~(\ref{eq:eccdecompose}) can be approximated as nearly monochromatic.}

We note that there have been many works studying the analytic waveform models for eccentric binaries \citep[see, e.g., ][for a review]{Moore2018,Moore_2019b,Loutrel+20, Knee2022ecc}. However, we still lack sufficient validation of the pN waveform against numerical relativity for the extreme eccentricity case (e.g., $e>0.999$). Despite this, the x-model can still give an accurate description of the highly eccentric waveform in the parameter space we consider in this paper. This argument is partly justified because we are focusing on the inspiral stage of stellar-mass binaries (i.e., the millihertz band for LISA detection) instead of the merger and ringdown stages in the LIGO band. For the systems that dominate the stellar-mass bursting source population in the local universe \citep{Xuan+23b,Xuan24bkg}, their pericenter distance is typically larger than $a(1-e)\gtrsim 10^{-3}{\rm au}$ (or in other words $\sim 10^{4}$ Schwarzschild radii). Therefore, even if the eccentricity is extreme, the gravitational field at the closest approach is still much weaker than the strength of the field for which the x-model has been validated against numerical relativity (21 cycles before the merger \citep{Hinder+10}), and the expansion serves as a plausible description of the GW signal for the systems we discuss hereafter.

Furthermore, due to the orbit of LISA around the Sun,
GW signals in the detector's output can have an annual modulation (see, e.g., Fig. 1 in Ref.~\citep{Xuan+23b}). Thus, we further include this effect by computing the LISA response function \citep{Cornish+03} (Note that, in general, LISA's annual motion is expected to enhance the angular resolution of sources' sky location measurement \citep[see, e.g.,][]{Cutler+98,Kocsis2007}):
\begin{equation}
\begin{aligned}
s_1(t) & =F^{+}(t) A_{+} \cos \left(2 \pi f\left[t+\hat{n} \cdot \mathbf{x}_1(t)\right]\right) \\
& +F^{\times}(t) A_{\times} \sin \left(2 \pi f\left[t+\hat{n} \cdot \mathbf{x}_1(t)\right]\right) \ ,\label{eq:response}
\end{aligned}
\end{equation}
where $s_1(t)$ represents one independent channel in the detector's output (Michelson response), $\hat{n}$ is the basis vector representing the sky location of the source, $\mathbf{x}_1(t)$ is the location of detector, $A_{+,\, \times}$ represents the strain amplitude for a monochromatic GW signal with frequency $f$, and $F^{+,\, \times}$ is the beam pattern factors:
\begin{figure}[htbp]
    \centering
    \includegraphics[width=3.5in]{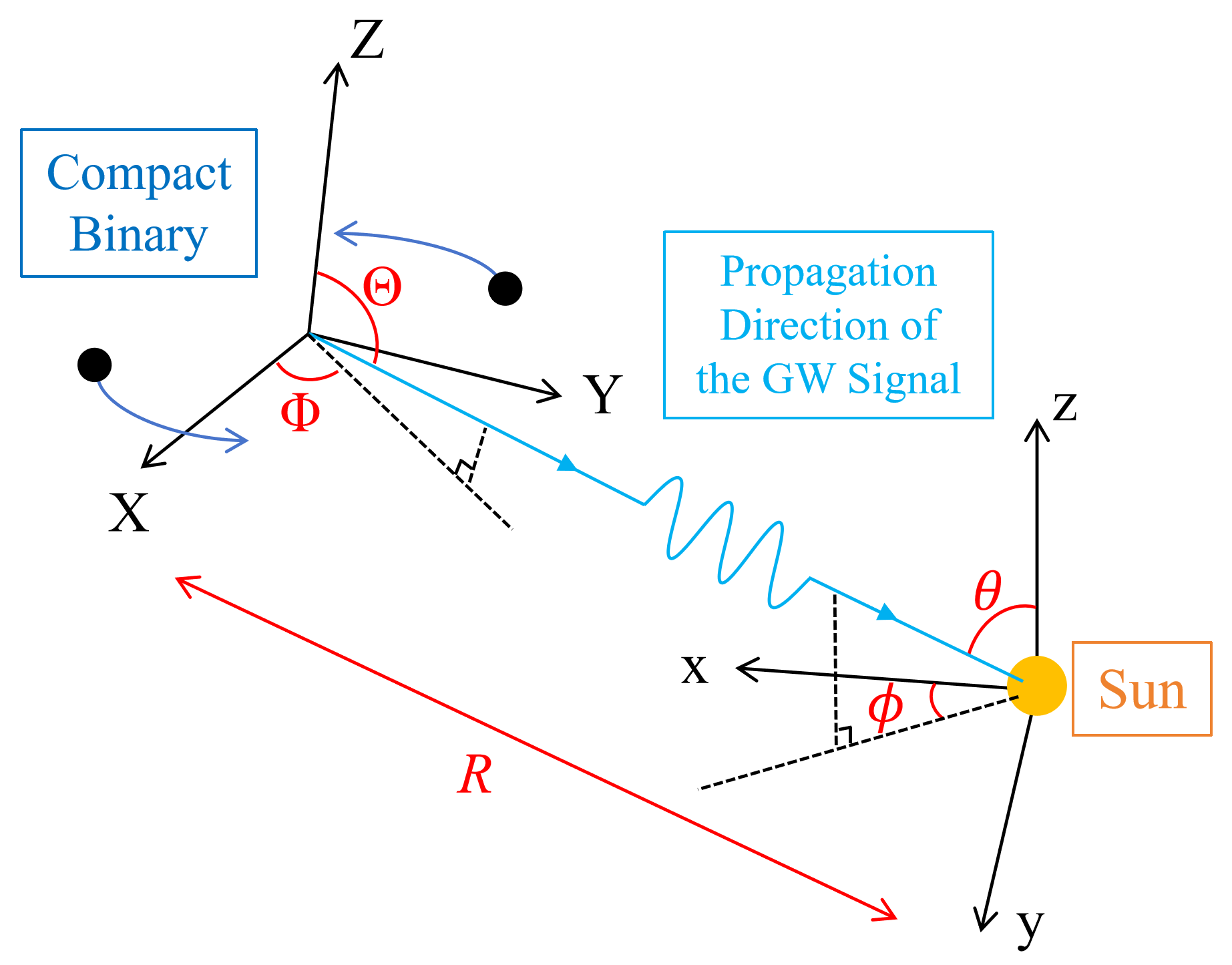} 
    \caption{{\bf The definition of $\Theta,\Phi,\theta,\phi$ and $R$ (not to scale).} Here we show an example compact object binary, which is emitting GW signal and being detected by LISA, to illustrate the definition of angular parameters in Section~\ref{sec:waveform_model}. The coordinate system $(X,Y,Z)$ is the non-rotating, comoving frame of the compact object binary, with the $Z$ axis parallel to the orbital angular momentum vector and the $X$ axis pointing in the direction of the (initial) pericenter. $\Theta,\Phi$ represent the spherical polar angles of the source's orientation relative to the observer (i.e., the propagation direction of the GW signal viewed in the source’s non-rotating, comoving frame). The coordinate system $(x,y,z)$ is the comoving frame of the Sun, with $\theta,\phi$ representing the sky location of the GW source. $R$ is the distance between the compact binary system and the Sun.
    }
    \label{fig:coordinate}
\end{figure}
\begin{equation}
\begin{aligned}
& F^{+}(t)=\frac{1}{2}\left(\cos 2 \psi D^{+}(t)-\sin 2 \psi D^{\times}(t)\right) \\
& F^{\times}(t)=\frac{1}{2}\left(\sin 2 \psi D^{+}(t)+\cos 2 \psi D^{\times}(t)\right),
\end{aligned}
\end{equation}
in which $\psi$ is the polarization angle of the GW signal (see Eq. 32 in \citep{Cornish+03}), and $D_{+,\,\times}(t; \theta, \phi)$ depends on the spherical polar angles of the GW source in the sky, $(\theta, \, \phi)$, as well as the detector's motion as a function of time (see Eqs. 54 -55 in Ref.~\citep{Cornish+03}). Note that the sky location angles $(\theta,\, \phi)$ are different from the intrinsic propagation direction of GW radiation viewed in the GW source's frame $(\Theta,\, \Phi)$, which we defined earlier in Equation~(\ref{eq:strains}).

\xzy{We note that the output of LISA can be considered like a pair of two-arm detectors, with two linearly independent signals (see, e.g., Sec.IIB2 in Ref.~\citep{Cutler+98} for more details). However, in this work, we only adopt the output of one channel, $s_1(t)$, for simplicity (see, e.g., Eq.46 in Ref.~\citep{Cornish+03}). This is due to the uncertainty of the realistic noise correlation between two channels and the complexity when considering multiple signals in data analysis. As a result, the detectability estimates presented here should be regarded as conservative, with the potential for improvement if the contributions from both channels are included in future studies.}

In Figure~~\ref{fig:coordinate}, we present an illustration of $\Theta,\Phi,\theta,\phi$ and $R$ definitions. In particular, $\Theta$ and $\Phi$ represent the propagation direction of the GW signal as viewed in the comoving frame of the compact object binary, with the binary's orbit lying in the $X-Y$ plane; $\theta$ and $\phi$ are the sky location of the GW source viewed in the comoving frame of the solar system, where the LISA detector undergoes annual motion around the sun. We note that the GW signal's polarization angle, $\psi$, is not explicitly shown in Figure~\ref{fig:coordinate}, which contributes to one more degree of freedom. In the calculation, $\psi$ can be worked out using $\psi=-\arctan (\hat{v} \cdot \mathbf{p} / \hat{u} \cdot \mathbf{p})$, where $\mathbf{p}$ is one of the principle polarization axes of GW beam, and $\hat{u}, \hat{v}$ represent two basis vectors in the $(x,y,z)$ coordinate system (see Eq.3, 31, 32 in Ref.~\citep{Cornish+03}).

Additionally, since the annual motion of detectors around the Sun has a frequency of $f_{\rm detector}=1\, yr^{-1}$, which is much lower than the frequency of GW signal in the millihertz band, we can take the adiabatic approximation. Under this approximation, the beam pattern factors, $F^{+,\, \times}$, remain nearly constant throughout each gravitational wave burst. Therefore, we can generalize Equation~(\ref{eq:response}) to the eccentric case by summing over contributions from each near-monochromatic harmonic (note that, for harmonic frequencies $f_n$ that predominantly contribute to the detection signal-to-noise ratio in the millihertz band, $F^{+,\, \times}$ can be treated as approximately constant and taken out of the summation \citep[see e.g.,][]{O'Leary+09,Wang+21,Xuan+23b}):
\begin{equation}
\begin{aligned}
s_1(t) & =F^{+}(t) \sum_{n=1}^{\infty}h_{n,\,+} \cos \left(2 \pi f_n\left[t+\hat{n} \cdot \mathbf{x}_1(t)\right]\right) \\
& +F^{\times}(t) \sum_{n=1}^{\infty}h_{n,\,\times} \sin \left(2 \pi f_n\left[t+\hat{n} \cdot \mathbf{x}_1(t)\right]\right) \\
& = F^{+}(t) h_{+}(t^{\prime}) +F^{\times}(t) h_{\times}(t^{\prime}) ,\label{eq:response2}
\end{aligned}
\end{equation}
where $h_{n,\,+}, h_{n,\,\times}$ represent the amplitude of the n-th harmonic in the eccentric GW signal (see Equation~(\ref{eq:eccdecompose})), $t^{\prime}=t+\hat{n} \cdot \mathbf{x}_1(t)$ is the delayed arrival time of GW signals caused by the motion of the detector.

As can be seen from Equation~(\ref{eq:strains}) - (\ref{eq:response2}), for a given antenna pattern of GW detector, the eccentric GW waveform in the detector's output can be parameterized using 10 parameters: the initial orbital parameters of the binary ($a_0,\,e_0$) (which enter the waveform via the initial condition), the total mass $M$, the symmetric mass ratio $\eta$ (or equivalently, the mass ratio $q=m_1/m_2$), the line of sight (propagation) direction of GW radiation in the source's frame ($\Theta,\, \Phi$), the location of GW source on the sky ($\theta,\, \phi$), the distance of GW source $R$, and the polarization angle $\psi$. 

Note that we set $\varphi(t=0)=0$ when introducing Equation~(\ref{eq:strains}); otherwise, there will be 11 parameters that affect the waveform, with the last one representing the initial orbital phase of the binary. Since this work focuses on the repeated burst sources, we assume that there is a significant number of bursts during observation. Thus, the initial phase of the first GW burst has little effect on the astrophysical interpretation, and we fix it for simplicity.

In the following sections, we will parameterize the GW templates using $\left\{f_{\rm orb,0}, 1-e_0, M, q, \cos\Theta, \Phi, \cos\theta,\phi, R,\psi \right\}$. Note that hereafter we use $f_{\rm orb}$ as an abbreviation for the initial orbital frequency $f_{\rm orb,0}$, which is related to $a_0$ via $f_{\rm orb,0}=(2\pi)^{-1}M^{1/2}a_0^{-3/2}$.

\section{Source identification and parameter extraction}\label{sec:totalfisherandMF}

\subsection{Fisher Matrix Analysis}
\label{sec:fisherana}

To evaluate the astrophysical information that can be extracted from highly eccentric GW sources, we adopt the Fisher matrix analysis \citep{Coe+09, Cutler+94}. This method is commonly used as a linearized estimation for the parameter measurement error in the high $\rm SNR$ limit.

We start by defining the noise-weighted inner product between two gravitational waveforms, $h_{1}(t)$ and $h_{2}(t)$: 
\begin{equation}
\left\langle h_{1} \mid h_{2}\right\rangle=2 \int_{0}^{\infty} \frac{\tilde{h}_{1}(f) \tilde{h}_{2}^{*}(f)+\tilde{h}_{1}^{*}(f) \tilde{h}_{2}(f)}{S_{\mathrm{n}}(f)} \mathrm{d} f \ ,
\label{eq:innerproduct}
\end{equation}
in which $\tilde{h}_l$ (with $l=1,2$) means a Fourier transform
of the waveform, and the star indicates the complex conjugate.

Representing the parameters of a GW source as a vector $\boldsymbol{\lambda}$, the GW waveform $h$ can be expressed as $h(t;\boldsymbol{\lambda})$. The Fisher matrix is defined as:
\begin{equation}
    F_{ij} = \left\langle\left.\frac{\partial h(\boldsymbol{\lambda})}{\partial \lambda_i}\right\rvert\, \frac{\partial h(\boldsymbol{\lambda})}{\partial \lambda_j}\right\rangle.\label{eq:fisherdefinition}
\end{equation}
in which $\lambda_i$ denotes the i-th parameter of the waveform.

We define $C$ as the inverse of the Fisher matrix, $C = F^{-1}$. It approximates the sample covariance matrix of the Bayesian posterior distribution for the GW source's parameters. In other words, we can use the following equation to estimate the error of parameter measurement:
\begin{equation}
\Delta \lambda_{i}= \sqrt{\left\langle\left(\delta \lambda_{i}\right)^{2}\right\rangle}=\sqrt{C_{i i}}\ .\label{eq:delt_estimation}
\end{equation}

\xzy{As shown in Equations~(\ref{eq:innerproduct})–(\ref{eq:delt_estimation}), the Fisher matrix is computed using the first-order derivatives of the waveform, without including higher-order corrections or Bayesian parameter estimation \citep[see, e.g.,][]{Christensen1998, Christensen2004}. This makes the Fisher matrix a fast and accessible tool in GW data analysis. However, as mentioned earlier in the Introduction, recent studies have shown that the Fisher matrix can give inaccurate results in certain regions of the parameter space when compared with full Bayesian parameter estimation (see, e.g., the comparison in Ref.~\citep{Toubiana_2020}, for testing General Relativity using stellar mass BBHs). Thus, we remind the reader that the results presented in this work are a preliminary estimate of the parameter measurement accuracy for bursting stellar mass BBHs. These estimates highlight the astrophysical potential of detecting bursting GW sources from LISA’s data stream. A full Bayesian analysis is left for future work.}


%

\subsubsection{Numerical Verification}\label{subsec:fishernum}
After getting Equations~(\ref{eq:innerproduct}) - (\ref{eq:delt_estimation}), we further combine them with the waveform model introduced in Section~\ref{sec:waveform_model} and estimate the parameter measurement error via numerical calculation. 

In particular, first, we generate the waveform template using the x-model (see Section~\ref{sec:waveform_model}) and parameterize the waveform as $h(t)=h(t;\boldsymbol{\lambda}=\left\{f_{\rm orb}, 1-e_0, M, q, \xzy{\cos\Theta}, \Phi, \xzy{\cos\theta},\phi, R,\psi \right\})$ (note that the detector's response function has been included in this step).

Second, we compute the partial derivative of the waveform relative to each parameter (see Equation~(\ref{eq:fisherdefinition})). For example, to get $\partial h/\partial M$, we vary the parameter $M\rightarrow M^{\prime}=M+d M$, then generate a new waveform $h^{\prime}(t)=h^{\prime}(t;\boldsymbol{\lambda}^{\prime}=\left\{f_{\rm orb}, 1-e_0, M+d M, q, \cos\Theta, \Phi, \cos\theta,\phi, R,\psi \right\})$. The partial derivative is obtained by computing the difference between $h^{\prime}$ and $h$, then divided by $d M$. 
\begin{figure*}[htbp]
    \centering
    \includegraphics[width=7.5in]{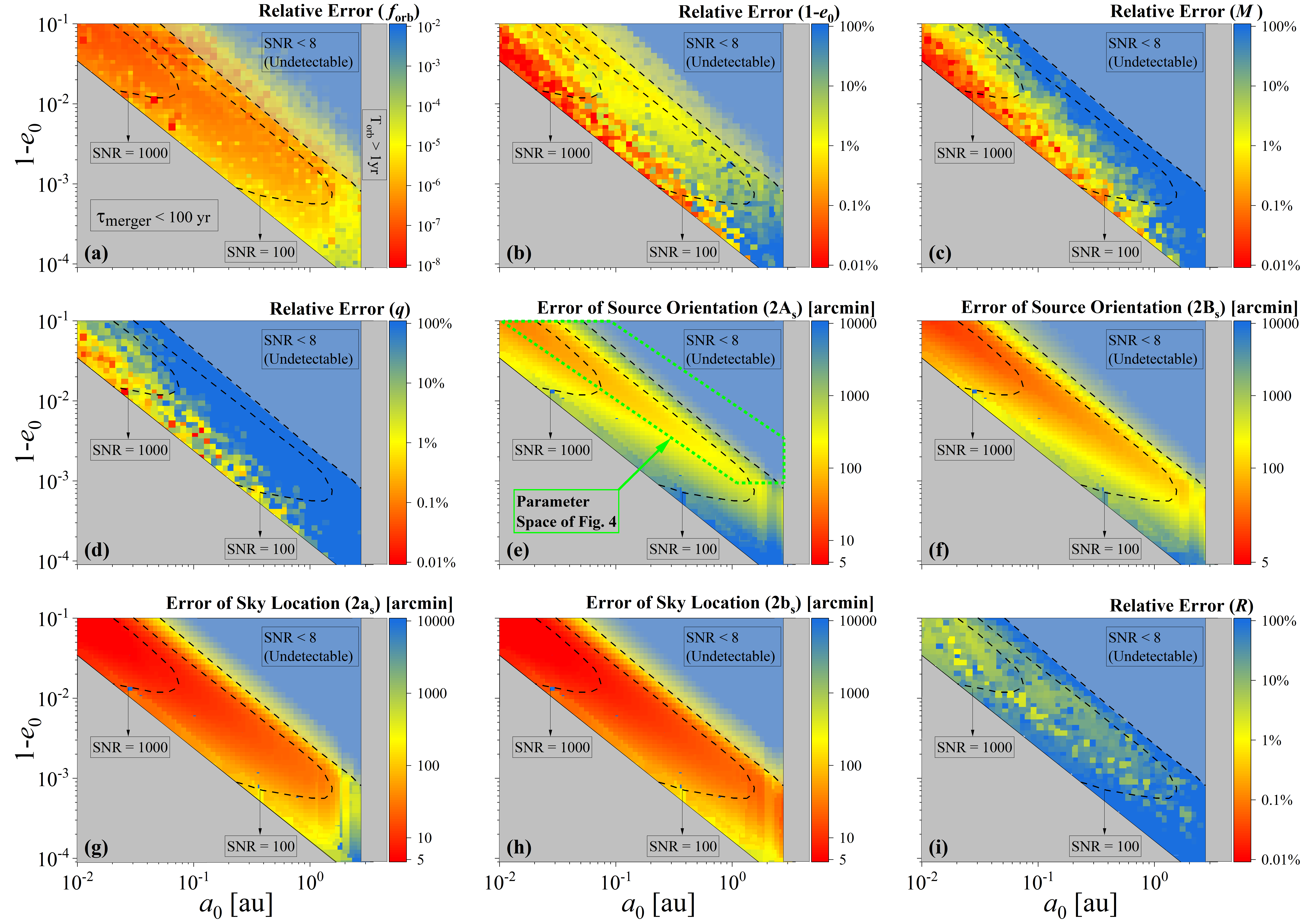} 
    \caption{{\bf{The dependence of a compact binary's parameter measurement error on its semi-major axis and eccentricity, computed using the Fisher matrix analysis (for 1-yr observation). }}
    Here we show a BBH system with $m_{1}=15$~M$_{\odot},$ $m_{2}=10$~M$_{\odot},$ (i.e., $M=25$~M$_{\odot},$ $q=3/2$), placed at
    $R=8$~kpc and observed for $1$~yr with LISA. In this example, we set $\Phi=\Theta=\theta=\phi=\pi/4$ for simplicity and map the compact binary's relative parameter measurement error as a function of its initial semi-major axis $a_0$ and eccentricity $1-e_0$ (see different colors in the figure). 
    In each panel, the dashed lines represent equal signal-to-noise ratio contours (analytically calculated using Eq. 17 in Ref.~\citep{Xuan+23b}), with $\rm SNR=8,\, 100,\, 1000$ from right to left. We exclude the parameter space where the binary has an orbital period $T_{\rm orb} > 1\,\rm yr$ (since their GW signal is not repeated bursts during the observation), or has a short merger timescale $\tau_{\rm merger} < 100\,\rm yr$ (since these short-living systems may have negligible number expectation in future detection). We note that the Fisher matrix analysis is computed for ten parameters ($f_{\rm orb}, 1-e_0, M, q, \cos\Theta, \Phi, \cos\theta,\phi, R,\psi$), while here we only show nine of them. This is because the last parameter $\psi$, which represents the polarization angle of the GW beam relative to the observer, is less relevant to the astrophysical inference, and we put its estimation in the appendix to avoid clutter. \xzy{Also, we convert the measurement accuracy of the source orientation, $\{\cos\Theta, \Phi\}$, and sky location, $\{\cos\theta, \phi\}$, into the major and minor axes of the sky error ellipsoids, $\{2A_s, 2B_s\}$, $\{2a_s, 2b_s\}$, respectively. See Appendix~\ref{app:skyerror} for more details.} Additionally, the green dotted lines in panel (e) of Figure~\ref{fig:fishernum} mark the parameter space that is later presented in Figure~\ref{fig:fishernum3}, for comparison.}
    
    \label{fig:fishernum}
\end{figure*}

We note that each partial derivative is a time series which represents the difference in the waveform caused by varying one of the parameters slightly around the central value. Thus, after getting all the partial derivative waveforms, we can adopt Equations~(\ref{eq:innerproduct}) and (\ref{eq:fisherdefinition}) to compute their inner products, get the Fisher matrix, then take the inverse of the matrix to estimate the parameter measurement error (see Equation~(\ref{eq:delt_estimation})).

\xzy{In this work, the parameter set, $\left\{f_{\rm orb}, 1-e_0, M, q, \xzy{\cos\Theta}, \Phi, \xzy{\cos\theta},\phi, R,\psi \right\}$, are intentionally chosen to avoid intrinsic degeneracies in the waveform model. However, since the aforementioned steps involve first-order-finite-difference derivatives and the numerical inversion of the Fisher matrix, they can still result in ill-conditioned outputs if not handled carefully (see, e.g., Sections IV and V in Ref.~\citep{Vallisneri_2008}). To ensure numerical stability, we choose the parameter variation $d\lambda_i$ such that $\left\langle\left.dh\right\rvert\, dh\right\rangle \sim 10^{-3} \left\langle\left.h\right\rvert\, h\right\rangle$. This choice keeps the change in the waveform small when computing the numerical derivative, maintaining the accuracy of the Fisher matrix analysis. Furthermore, we rescale the unit of source parameters to improve the condition number and reduce the magnitude gap between the largest- and smallest-modulus eigenvalues. When computing the Fisher matrix for a given point, we design the code to automatically try and adjust the aforementioned properties (i.e., the finite variation in numerical derivative and the rescaling of parameters), to keep the resultant Fisher matrix non-singular.}

In Figure~\ref{fig:fishernum}, we show the Fisher matrix analysis results for the GW signal from a BBH system with $m_1=15\, {\rm M_{\odot}}$, $m_2=10\, {\rm M_{\odot}}$, placed at $R=8\rm kpc$, with the position of $\Phi=\Theta=\theta=\phi=\pi/4$, and observed for $T_{\rm obs}= 1\, \rm yr$. In particular, we generate different initial conditions of the system by changing $a_0$ and $e_0$. For each configuration, we compute the partial derivative of the waveform relative to the aforementioned 10 parameters ($f_{\rm orb}, 1-e_0, M, q, \cos\Theta, \Phi, \cos\theta,\phi, R,\psi$), estimate their Fisher matrix, and get the measurement error. We plot the relative error of each parameter in different colors and map the result as a function of $(a_0,1-e_0)$. \xzy{To better compare with observation, we convert the measurement accuracy of the source orientation, $\{\cos\Theta, \Phi\}$, and sky location, $\{\cos\theta, \phi\}$, into the major and minor axes of the sky error ellipsoids, $\{2A_s, 2B_s\}$, $\{2a_s, 2b_s\}$, respectively. See Appendix~\ref{app:skyerror} for more details.}

Note that this system serves as a representative example of the parameter measurement accuracy (see, e.g., the characteristic parameters of bursting BBHs in the population analysis of Ref.~\citep{Xuan+23b}). In realistic observations, however, the GW sources can have different orientations and sky locations, which can potentially affect the parameter measurement error (note that it is also straightforward to analyze other systems with different values of $(\Phi,\Theta,\theta,\phi)$). We emphasize that there have been previous works estimating the change in parameter measurement errors as a function of different initial conditions of the system (see, e.g., studies on supermassive black hole binaries \citep{Kocsis2007,Kocsis_2008,Mikoczi+12}). In our case, the width of the distribution of parameter estimation errors could be significantly narrower, as the bursting BBHs are observed well before their merger, with slower orbital evolution. This provides a more stable signal, which allows for more precise parameter extraction. Furthermore, this work focuses on the stellar mass bursting BBHs in the local universe, but a similar approach may be applied to other cases, such as the parameter estimation of highly eccentric extreme mass-ratio inspirals (EMRIs) at a cosmological distance \citep[see, e.g.,][]{barack04,Rubbo06,Yunes_2008EMRIburst,Berry_2013,Han+20EMRIburst,oliver2023gravitational,Fan22,Naoz+23}. 

Furthermore, Figure~\ref{fig:fishernum} covers the parameter space where bursting stellar-mass BBHs are most likely to be detected in the Milky Way with future millihertz GW detectors \citep{Xuan+23b,Xuan24bkg}. In particular, as discussed in Section~\ref{section:intro}, many dynamical channels can generate wide, highly eccentric BBH systems. During the evolution, if these systems enter the parameter space with an orbital period smaller than the observation time (i.e., to the left of the grey-colored region with $T_{\rm orb} > 1\,\rm yr$), and a signal-to-noise ratio larger than the detection threshold (i.e., below the dashed line of $\rm SNR=8$), their GW signal will be identified as ``repeated bursts", and can be used for parameter extraction. Additionally, we calculate the Fisher matrix only for systems with $\tau_{\rm merger}>100\,\rm yr$ (to the top right of the grey-colored triangle). This is because the sources to the top right of each panel will have a longer merger timescale compared with those on the bottom left (see, e.g., Fig.4 in Ref.~\citep{Xuan+23b}), thus are more likely to be detected due to their extended detectable time in mHz band (note that this also indicates the relative error estimation of sources with moderate SNR is more representative of the accuracy we will get in future observation of stellar-mass bursting sources, since they have the longest lifetime and largest number expectation among all the detectable systems). On the other hand, the expected number of short-living systems with $\tau_{\rm merger}<100\,\rm yr$ is negligible (see, e.g., Section 3.2 in Ref.~\citep{Xuan+23b}), and the computational expense of their Fisher matrix is large, so we decide not to include them in this example.

\begin{figure*}[htbp]
    \centering
    \includegraphics[width=7in]{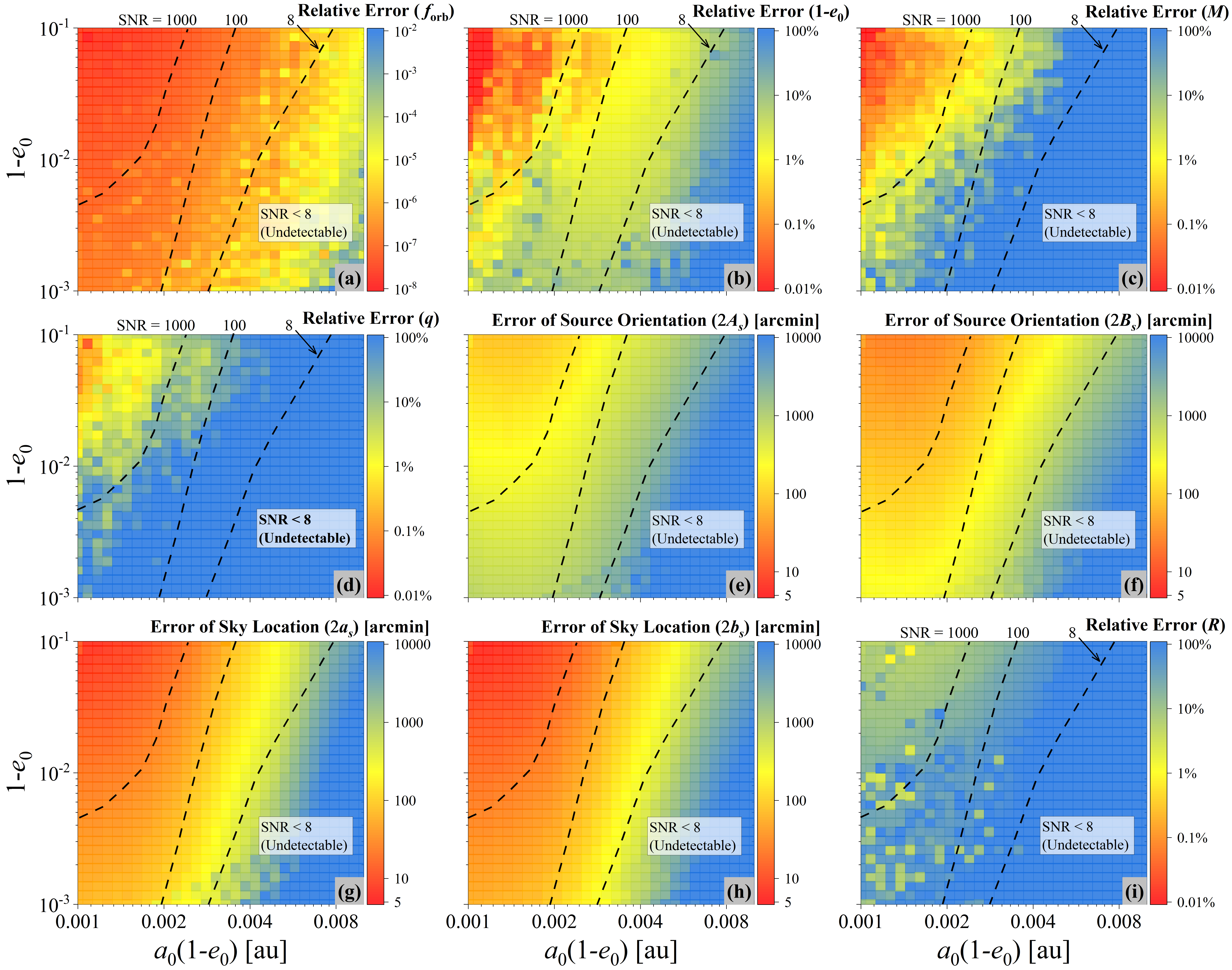} 
    \caption{{\bf{The dependence of a compact binary's parameter measurement error on its initial pericenter distance, $r_p= a_0(1-e_0)$, and eccentricity $e_0$ (for 5-yr observation)}.  }
    Here, we consider the same systems as in Figure~\ref{fig:fishernum}, but re-parameterize the x-axis using $r_p= a_0(1-e_0)$ and zoom in to highlight the parameter space where we expect to detect most of the stellar mass bursting BBHs \citep{Xuan+23b,Xuan24bkg}, i.e., $
    \rm SNR\sim 8 - 1000$. Additionally, we changed the observation time from 1 year to 5 years to show an optimistic estimation of measurement accuracy (for the 1-year case, see Figure~\ref{fig:fishernum2} in Appendix), and mark the parameter space of this figure with green dotted lines in panel (e) of Figure~\ref{fig:fishernum} for comparison purposes.}
    \label{fig:fishernum3}
\end{figure*}

Additionally, to provide an optimistic estimation of the measurement accuracy, we extend the observation time to 5 years and repeat the analysis from Section~\ref{subsec:fishernum}. For our analysis, we focus on the parameter space highlighted by the green dotted line in panel (e) in Figure \ref{fig:fishernum}. The results are summarized in Figure~\ref{fig:fishernum3}, where we consider the same systems as in Figure~\ref{fig:fishernum}, but parameterized the x-axis using the pericenter distance of the binary system, $r_p= a_0(1-e_0)$. As shown in Figure~\ref{fig:fishernum3}, this choice of x-axis ($r_p$), along with a zoomed-in view of the parameter space, highlights the region where a bursting BBH system has a signal-to-noise ratio of $\rm SNR\sim 8 - 1000$, and provides more detailed estimation results. 

We emphasize that, Figures~\ref{fig:fishernum} and \ref{fig:fishernum3} show the parameter measurement accuracy for bursting sources in the Milky Way (at a representative distance of $8$~kpc). However, these results can be generalized to
bursting sources in the local universe, provided the redshift is negligible. In particular, if a compact binary is located at a distance $R$ different from $8$~kpc, the overall amplitude of its GW signal will scale inversely with the distance ($h\sim R^{-1}$, see Equation~(\ref{eq:strains})). Furthermore, because the Fisher matrix elements are proportional to the square of the waveform amplitude ($F_{ij}\sim h^2\sim R^{-2}$, see Equation~(\ref{eq:fisherdefinition})), the parameter measurement error, $\Delta \lambda_i$, after inverting the Fisher matrix and taking the square root, should scale proportionally to the distance ($\Delta \lambda_{i}\sim C_{ii}^{\frac{1}{2}}\sim R$, see Equation~(\ref{eq:delt_estimation})). In other words, the results of Figures~\ref{fig:fishernum} and \ref{fig:fishernum3} can be rescaled using the following relation:
\begin{equation}
    \Delta \lambda_i^{\prime}(R)\sim \Delta \lambda_i(8\,{\rm kpc}) \left( \frac{R}{8\,{\rm kpc}}\right) \ ,
\end{equation}
where $\Delta \lambda_i^{\prime}(R)$ represents the parameter measurement error for highly eccentric systems at a distance $R$, and $\Delta \lambda_i(8~{\rm kpc})$ refers to the Fisher matrix analysis result (absolute error) for systems at $8~{\rm kpc}$, which can be estimated using Figures~\ref{fig:fishernum} and \ref{fig:fishernum3}.

\subsubsection{Astrophysical Interpretation}
\label{subsec:astrointerpretation}
In this section, we will discuss the astrophysical interpretation of each panel (system's parameter) in Figures ~\ref{fig:fishernum} and ~\ref{fig:fishernum3}. First, as shown in panel (a) of these two figures, the relative error ($\Delta f_{\rm orb}/f_{\rm orb}$) of orbital frequency, $f_{\rm orb}$, is smaller than $\sim 10^{-6}$ for most of the detectable cases. In other words, considering the value of $f_{\rm orb}$ for a $10-15\rm M_{\odot}$ BBH system with semi-major axis $a\sim 10^{-2} - 1\rm au$, the absolute error of orbital frequency measurement is roughly $\Delta f_{\rm orb}\sim 10^{-12} - 10^{-10}\, \rm Hz$ in the parameter space we consider. 

Such high accuracy can be understood analytically. For example, the measurement accuracy depends on the waveform's sensitivity to changes in $f_{\rm orb}$. Therefore, we can vary $f_{\rm orb}$ by $d f_{\rm orb}$, then identify the value of $d f_{\rm orb}$ at which the new waveform significantly differs from the original one. Thus, this critical value is used to estimate the measurement accuracy, $d f_{\rm orb}\sim \Delta f_{\rm orb}$. The concept here is similar to the Fisher matrix analysis (see Equations~(\ref{eq:fisherdefinition}) and (\ref{eq:delt_estimation}). Note that $F_{ij}$ in the Fisher matrix analysis is, in general, estimating the waveform's sensitivity to the changes of parameters.). 



Following the aforementioned method, we consider two GW waveforms with slightly different orbital frequencies, i.e., $h_1(t;f_{\rm orb})$ and $h_2(t;f^{\prime}_{\rm orb}=f_{\rm orb}+d f_{\rm orb})$. Specifically, for a highly eccentric binary, the GW waveform can be approximated by a series of bursts, separated by $T_{\rm orb}$ in time. Because of the difference in orbital frequency, $h_2$ will have a different period of GW burst compared to $h_1$, which can be described using $d T_{\rm orb}=d(1/f_{\rm orb})\sim d f_{\rm orb}/f_{\rm orb}^2$ (note that we assume $f_{\rm orb}$ and $f^{\prime}_{\rm orb}$ are constants during the observation, i.e., the sources chirp slowly).
Therefore, the position of each GW burst in $h_2$ will shift in time relative to the bursts in $h_1$. During an observation time of $T_{\rm obs}$, the accumulated difference (i.e., position shift of the arrival time for the last GW burst of $h_2$, compared with $h_1$) can be estimated by:
\begin{equation}
   d t_{\rm last\, burst}  \sim N\cdot d T_{\rm orb} = T_{\rm obs}f_{\rm orb} \cdot \frac{d f_{\rm orb}}{f_{\rm orb}^2},\label{eq:dt lastburst}
\end{equation}
where $N$ represents the number of bursts detected during the observation.

In realistic observation, two different GW waveforms can be distinguished when their match is below a given threshold (see, e.g., Eq.~28 in Ref.~\citep{Xuan+21}). For our cases, however, we simplify the condition and assume that $h_1$ is distinguishable from $h_2$ when the position shift of the last GW burst, $d t_{\rm last burst}$, is greater than the width of that GW burst, $T_{\rm burst}$.
\begin{equation}
   d t_{\rm last burst}>T_{\rm burst}.\label{eq:mismatch condition}
\end{equation}
This estimation is partly justified because the time domain overlap between two waveforms, $\int h_1(t)h_2(t) dt$, will drop to $\lesssim 1/2$ of the fully matched value when Equation~(\ref{eq:mismatch condition}) is satisfied, which significantly reduces the match between $h_1,\, h_2$ and potentially makes them distinguishable in data analysis.

In Equation~(\ref{eq:mismatch condition}), the width of a GW burst, $T_{\rm burst}$, can be estimated using the binary's pericenter passage time, $T_p$, \citep[see, e.g.,][]{O'Leary+09}: 
\begin{equation}
   T_{\rm burst}\sim T_{\rm p} \sim  (1-e)^{3/2} T_{\rm orb} \ .
    \label{eq:tburst}
\end{equation}
Thus, we can plug in Equations~(\ref{eq:dt lastburst}) and (\ref{eq:tburst}) into Equation~(\ref{eq:mismatch condition}), and get a heuristic estimation of the measurement accuracy in orbital frequency, $\Delta f_{\rm orb}$:
\begin{equation}
   \Delta f_{\rm orb}\sim d f_{\rm orb}=\frac{f_{\rm orb}}{T_{\rm obs}}d t_{\rm last burst} > \frac{f_{\rm orb}}{T_{\rm obs}}T_{\rm burst} \sim \frac{(1-e)^{\frac{3}{2}}}{T_{\rm obs}} \ .\label{eq:analyticalestiforb}
\end{equation}
In other words, our waveform analysis will be sensitive to any change in $f_{\rm orb}$ greater than $\sim(1-e)^\frac{3}{2}/T_{\rm obs}$. For a 1-yr observation of a BBH system with $(1-e)\sim 10^{-3}-10^{-1}$, Equation~(\ref{eq:analyticalestiforb}) yields $\Delta f_{\rm orb}\sim 10^{-12}-10^{-9} \rm Hz$, which is consistent with the numerical result we discussed at the beginning of this section (see, e.g., Figure~\ref{fig:fisherana} in Appendix for more details).

We note that Equation~(\ref{eq:analyticalestiforb}) only serves as a heuristic estimation of $\Delta f_{\rm orb}$. It neglects the GW source's orbital frequency shift during observation, which limits the application of this formula in the high $\rm SNR$ region, where the binary quickly loses energy and shrinks. However, for the majority of stellar-mass binaries we consider in this work, especially those close to the dashed line of $\rm SNR=8$ in Figure~\ref{fig:fishernum}, the GW merger timescale is typically much longer than the observation time (see, e.g., Fig.4 in Ref.~\citep{Xuan+23b}). In this cases, Equation~(\ref{eq:analyticalestiforb}) yields a quick analytical estimation of $f_{\rm orb}$ measurement error, and can help with determining the grid size of matched filtering, which we will introduce soon in the next section.

Second, the eccentricity measurement accuracy reaches $\Delta(1-e_0)/(1-e_0)\sim 1\%$ for most detectable cases (see panel (b) of Figure~\ref{fig:fishernum}), allowing for precise constraints on the binary's dynamical origin (note that different dynamical channels may yield different characteristic eccentricities in the same frequency band, see, e.g., Fig.4 of Ref.~\citep{Martinez+20}).  Moreover, for 5-year integration, the accuracy may reach below $0.1\%$ for a significant part of the parameter space, as depicted in Figure~\ref{fig:fishernum3}. This accuracy highlights the dominant role of eccentricity in shaping the overall form of the bursting GW waveform, as the term $1-e$ is directly related to the width of each GW burst we observe (see Equation~(\ref{eq:tburst})), and the waveform fitting is thus very sensitive to eccentricity.

Third, as illustrated in panels (c) and (d) of Figure~\ref{fig:fishernum}, mHz GW detection may be less sensitive to the total mass $M$ and mass ratio $q$ of a bursting binary system. For example, during a 1-year observation, we can only constrain the total mass of bursting stellar mass BBHs ($\Delta M/M<10\%$) when their $\rm SNR$ exceeds $\sim 100$. Moreover, for determining mass ratio $q$, the bursting sources need to have $\rm SNR$ greater than $\sim 1000$. This result is consistent with our analysis in Section~\ref{sec:astro}. In particular, bursting sources have small radiation power compared to moderately eccentric sources in the same frequency band. Therefore, their orbital evolution is slow, and frequency shift rate $df_{\rm orb}/dt$ is hard to measure, which limits the accuracy of mass measurement \citep[see, e.g.,][]{Cutler+94}.

However, the measurement accuracy for $M$ and $q$ can significantly improve with extended observation time. For example, as shown in panel (c) of Figures~\ref{fig:fishernum3} and \ref{fig:fishernum2} (in appendix), when increasing observation time from 1 year to 5 years, the total mass measurement accuracy is enhanced by a factor of $\sim 10$, which allows us to measure the mass of bursting binary system in many of the detectable cases (with moderate $\rm SNR$). In other words, for the first $\sim 1$ year of LISA observation, we may not get a good constraint on the mass of stellar-mass bursting sources with $\rm SNR \lesssim 100$, but most of them will have a mass measurement by the end of 5-year LISA mission (except for some marginal detectable cases with $\rm SNR\sim 8$ for 5 years), which enables us to distinguish between different bursting sources, such as highly eccentric BBHs, binary neutron stars (BNSs), and double white dwarfs (DWDs). 

\xzy{Furthermore, as shown in panels (e) - (h) of Figures~\ref{fig:fishernum} and \ref{fig:fishernum3}, the orientation (as the axes of error ellipsoid, $\{2A_S, 2B_s\}$, \citep{Lang2006,Kocsis_2008,Kocsis2007,Mikoczi+12}) of bursting binary systems can only be constrained to an accuracy of $\sim 100-1000 $~arc minutes when SNR exceeds $\sim 100$ (similar for 1-year or longer integration times). However, our analysis suggests that the detection will be more sensitive to the sky location of the sources. In particular, for a 5-year observation, the resulting sky error ellipsoids exhibit comparable angular sizes for $a_s$ and $b_s$, typically ranging from $\sim 5 - 500$ arc minutes, depending on the $\rm SNR$. \xzy{This sky localization can enable the identification of specific host environments of GW sources in the Milky Way, such as globular clusters, the Galactic center, and the Galactic field, which potentially helps us to disentangle different formation channels of BBHs.}}


As shown in panel (i) of Figures~\ref{fig:fishernum} and \ref{fig:fishernum3}, the distance of a bursting binary system is only marginally constrained in many detectable cases. For example, during a 1-year observation, $\Delta R/R$ reaches $\sim 10\%$ only when $\rm SNR$ exceeds $\sim 100$. This accuracy improves when we extend the observation time to 5 years, but the relative error of $R$ still remains $\gtrsim 10\%$ for most of the systems with $8<{\rm SNR} <100$. Such low accuracy of distance measurement is also caused by the slow orbital evolution of the bursting binary system (see e.g., a similar discussion in the explanation of panels (c) and (d)). In particular, for bursting sources at low redshift, $R$ only affects the overall amplitude of the GW signal (see the term $M\eta/R$ in Equation~(\ref{eq:strains})). Therefore, if the GW source undergoes slow orbital evolution, the effect of $R$ will degenerate with other mass parameters (such as the total mass $M$), thus adding to the difficulty of parameter extraction. However, we may still get a heuristic estimation of the binary's formation environment using the distance measurement, since an accuracy of $\sim 10\%$ is good enough to infer the binary's location in the local universe. Furthermore, with the potential for electromagnetic counterparts (e.g., if the source resides in a known stellar population like a globular cluster or contains a directly observable white dwarf), we may be able to disentangle the distance from other orbital parameters, thereby improving constraints on the location of the bursting sources.


We note that, in Figures~\ref{fig:fishernum} and \ref{fig:fishernum3}, there are some data points with a relative error estimation inconsistent with the general distribution in the surrounding region (see, e.g., some of the bright-yellow points, to the bottom-left in panel (e) of Figure~\ref{fig:fishernum3}). Such fluctuation is caused by numerical error and degeneracy in the waveform model. In particular, for some bursting binary systems, only a few to a few dozen GW bursts will be detected during an observation time of $\sim 1-5$ years, which limits the numerical accuracy in the frequency domain inner product of waveforms. Furthermore, among all the 10 parameters we discuss (see Section~\ref{sec:waveform_model}), many of them will have a similar effect on the overall shape of the GW signal (e.g., the aforementioned degeneracy between distance $R$ and total mass $M$). Therefore, with a certain combination of parameter values, the Fisher matrix may become singular in the numerical analysis, which leads to inconsistent relative error estimation.

Additionally, we emphasize that, it is possible to analytically estimate the parameter measurement error, especially in the cases when the GW waveform model is simple (see, e.g., Eqs.~27 - 29 in Ref.~\citep{Seto02WD}). In our cases, however, the complexity of the bursting system's parameters makes it hard to simplify the cross-terms in the Fisher matrix via an analytical approach. Therefore, although we tried to develop an analytical estimation, the outcome did not yield a well-performed constraint on the measurement accuracy, as shown in the Appendix~\ref{appendix:fisher}. However, for the parameters that play a dominant role in the overall shape of bursting GW signal, such as the orbital frequency $f_{\rm orb}$, the analytical method can still be accurate even if we make many approximations (see Equation~(\ref{eq:analyticalestiforb})). Therefore, it can yield an efficient estimation with no need to compute the numerical GW waveform.

\subsection{Matched Filtering}\label{sec:MFmethod}
In Section~\ref{sec:fisherana}, we discussed the parameter measurement accuracy of highly eccentric GW sources. However, to get such astrophysical information in realistic observation, we need to identify the bursting GW signals and extract them from the data stream. Therefore, in this section, we will briefly introduce the technique of matched filtering, which is commonly adopted in GW data analysis to detect signals from astrophysical sources \citep[see, e.g., ][for more details]{Cutler+94, Allen2011,Petigura2013MF,LIGO2021SoftX..1300658A,LIGO2023tutorial}.

In particular, the measured strain amplitude in the detector's output can be described using 
$s(t) = h(t)+ n(t)$, with the (possibly present) signal $h(t)$, and the detector noise $n(t)$. To extract the targeted GW signal, we first generate the GW waveform template based on our understanding of astrophysical sources, $h_{\rm template}(t)$, then convolve $s(t)$ with $h_{\rm template}(t)$ in the frequency domain \citep[see, e.g.,  Eq.4.1 in][]{Allen2011}:
\begin{equation}
\begin{aligned}
x\left(t_0\right) & =2 \int_{-\infty}^{\infty} \frac{\tilde{s}(f) \tilde{h}_{\text {template }}^*(f)}{S_n(f)} d f,\label{eq:MF}
\end{aligned}
\end{equation}
in which $\tilde{s}(f)$ and $\tilde{h}_{\rm template }(f)$ are the Fourier transform of $s(t)$ and $h_{\rm template}(t)$, and the star stands for the complex conjugate. We note that, for GW mergers, the parameter $t_0$ represents the termination time (the time at the detector at which the coalescence occurs). Furthermore, we can change the termination time of the template $h_{\rm template}(t)$ and find the value of $t_0$ when the norm of $x(t_0)$ is maximized, i.e., $|x(t_0)|_{\rm max}=|x_m|$. Thus, this $t_0$ represents the correct (or the most likely) location of the GW signal in the detector's output. In our cases, the bursting binary may not merge for the entire observation. However, we can still use $t_0$ (with a modified definition) to describe the position shift of $h_{\rm template}(t)$ in the time domain. Hereafter, we define $t_0$ as the time at which the first GW burst of $h_{\rm template}(t)$ occurs.

As shown in Equation~(\ref{eq:MF}), for a given template $h_{\rm template}$, we can find the maximum norm of the convolution result, $|x_m|$. This quantity, with a proper normalization constant $\sigma_m$ (see, e.g., Eq.~4.3 in Ref.~\citep{Allen2011}), represents the optimized signal-to-noise ratio for the template we use in the search, \begin{equation}\label{eq:SNRm}
    \rm SNR_{m}=\frac{|x_m|}{\sigma_m} \ ,
\end{equation} 
note that the amplitude of $h_{\rm template}(t)$ will cancel in the computation of $\rm SNR_{m}=|x_m|/\sigma_m$, thus we can place the GW source at an arbitrary distance $R$ when constructing the template. In realistic detection, if this signal-to-noise ratio $\rm SNR_{m}$ exceeds a certain threshold, we will identify the corresponding template $h_{\rm template}(t)$ as existing in the detector's output $s(t)$. In other words, we successfully identify the GW source and measure its parameters using the parameters of $h_{\rm template}$. 

In Section~\ref{sec:mockobservation}, we use the aforementioned method to conduct matched filtering and test the detectability of bursting sources in an example mock LISA observation. Additionally, in this work, we use the software package PyCBC for the numerical realization \citep{PYCBC_alex_nitz_2024_10473621}, and acknowledge the supporting materials provided by the Gravitational Wave Open Science Center of the LIGO Scientific Collaboration, the Virgo Collaboration, and KAGRA \citep{AlvinLi2023,LIGO2021SoftX..1300658A,LIGO2023tutorial}.


\section{An Example of Mock Data Analysis}
\label{sec:mockobservation}

\subsection{The Difference Between a Bursting and Non-Bursting Data Analysis}\label{sunsec:differenceofburstfitting}
As shown in Section~\ref{sec:totalfisherandMF}, highly eccentric compact binaries can provide us with valuable information about their orbital parameters and formation environment. However, since their GW signals are made up of transient bursts in the time domain, the frequency power spectrum of bursting GW waveform will thus have many ($\sim 10^3-10^6$) harmonics that contribute significantly to the total signal-to-noise ratio \citep[see, e.g., ][]{Kocsis_2012,Xuan+23b}. Such a large number of harmonics distributes the energy of the GW signal in a wide range of frequencies, thus potentially adding to the difficulty of realistic data analysis. In particular, many existing data analysis methods, such as matched filtering, rely on accurate template fitting in the frequency domain; however, the bursting waveform could have many of the harmonics out of the sensitive frequency band of LISA. Furthermore, since there could be a large population of bursting sources, their collective GW signal may form a confusion background, with many GW harmonics from various sources overlapping with each other \citep{Xuan+23b}. 

Therefore, in this section, we will explore whether these bursting binaries could be efficiently distinguished when multiple sources and instrumental noises are present simultaneously in the detector's output. Specifically, we carry out mock LISA observation, generating a population of wide, highly eccentric BBHs based on our previous studies of the Galactic center \citep{Xuan+23b}. We then compute the collective GW signal from these bursting BBHs, combine the signal with the detector's instrumental noise and astrophysical foreground caused by unresolved Galactic binaries \citep{Cornish2017galacticforeground,Robson+19LISAnoise}, and apply the matched filtering as described in Section~\ref{sec:MFmethod} to extract the information of the bursting BBHs. 

\subsection{Population Model and Waveform Generation}\label{subsec: population}

As a proof of concept, we consider the bursting BBHs formed in the Galactic Center as a representative example for the mock observation. In particular, the Milky Way's Galactic nucleus offers a promising environment for the formation of wide, highly eccentric BBHs \citep[see, e.g.,][]{Kocsis_2012,Hoang+18,Hoang+19,Stephan+19,Wang+21,Rose+23collisiongc,Arca+23,Zhang24}. For example, the orbital eccentricity of BBHs can be excited by the supermassive black hole in the galactic nucleus because of the Eccentric Kozai-Lidov mechanism \citep{Kozai1962,Lidov1962,Naoz16}, resulting in the bursting signatures on their GW signal. In our previous works \citep{Hoang+18,Xuan+23b,Xuan24bkg}, we have carried out detailed simulations of the aforementioned systems, including the secular equations (for the hierarchical triple system) up to the octupole level of approximation \citep{Naoz+13}, general relativity precession \citep[e.g.,][]{naoz13}, and GW
emission \citep{Peters64,Zwick+20}. Thus, here, we adopt the result of the simulation presented in Ref.~\citep{Xuan+23b} for the bursting BBH population in the mock LISA data.

\begin{figure}[htbp]
    \centering
    \includegraphics[width=3.5in]{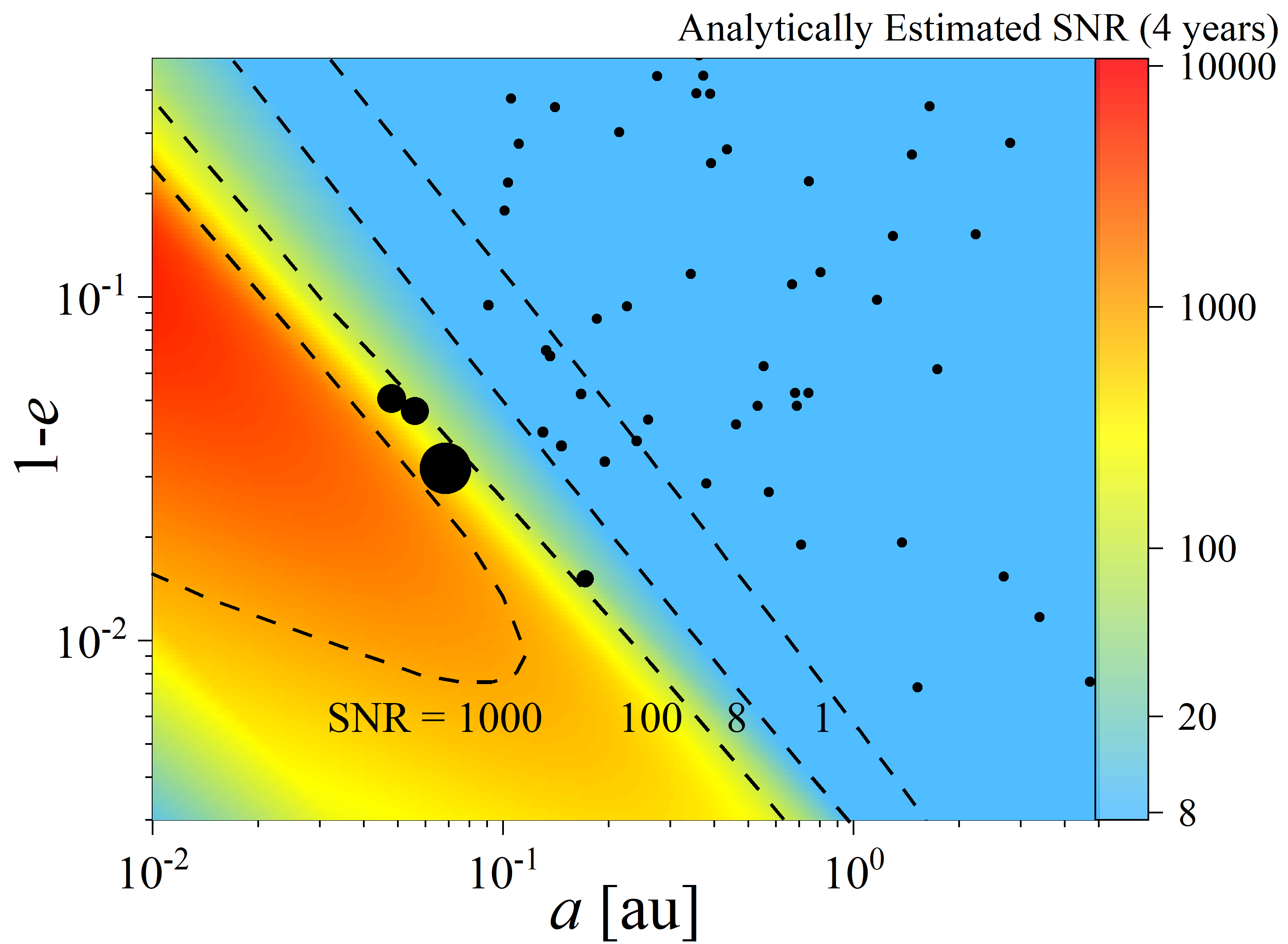} 
    \caption{{\bf An example of the intrinsic parameters of highly eccentric BBHs used in our mock LISA observation.} Here, we adopt the simulation result of the compact binary population in the Milky Way's nuclear star cluster (see Section 3.5 in Ref.~\citep[][]{Xuan+23b}) and show the parameters of highly eccentric BBHs in black dots. The x-axis represents the binary's semi-major axis $a$, and the y-axis represents one minus the binary's eccentricity $(1-e)$. We estimate the overall signal-to-noise ratio of the BBH system during a 4-yr LISA observation, depicted in different colors. The dashed lines represent equal signal-to-noise ratio contours for SNR = 1000, 100, 8, and 1, from left to right. Here, the $\rm SNR$ is analytically calculated using Equation~(\ref{eq:SNR}), assuming $m_{1}=m_{2}=10$~M$_{\odot}$ and $R=8$~kpc for simplicity. However, we note that the mass of BBHs in the mock data analysis is generated from the stellar evolution model and may slightly differ from $\sim 10$~M$_{\odot}$. Therefore, the numerical SNR is expected to vary slightly from the analytical calculation.  For demonstration purposes, we enlarge the size of dots for the detectable BBH systems based on the SNR. 
    }
    \label{fig:population}
\end{figure}

Figure~\ref{fig:population} illustrates the intrinsic parameters of BBHs we get from the simulation mentioned above, with each black dot representing the orbital parameter $(a, 1-e)$ of a binary system. As shown in the Figure, there could be $\sim 50$ highly eccentric BBHs simultaneously in the nuclear star cluster at the Milky Way Galactic center (note that there are many BBHs with moderate eccentricity as well, which are not shown in Figure~\ref{fig:population}). Furthermore, in this sample, we find 4 BBHs with the expected signal-to-noise ratio exceeding the detection threshold (i.e., analytical $\rm SNR>8$  for a 4-year observation, estimated using Equation~(\ref{eq:SNR})). These systems, if properly identified, will become detectable GW sources. The mass of these BBH systems is determined using the stellar evolution model in Appendix A.3 of Ref~\citep{Xuan+23b}, with most of the black holes having a mass $\sim 10~{\rm M_{\rm \odot}}$. Additionally, the BBHs' orientation is generated randomly following an isotropic distribution, and we set their distance to the detector as $R=8~\rm kpc$.

Based on the aforementioned population model, we generate the time domain GW signal for each BBH system and compute the detector's response correspondingly (see Equations~(\ref{eq:strains}) - (\ref{eq:response2})). We then calculate the mock LISA data by combining all the BBHs' signals and adding the detector noise. For simplicity, we assume stationary Gaussian noise, and adopt the LISA noise spectrum from Ref.~\citep{Robson+19LISAnoise}, which includes the instrumental noise as well as the foreground from unresolved galactic binaries
\citep[see, e.g., ][]{Cornish2017galacticforeground}.

\subsection{Matched Filtering and Mock Observation Results}\label{subsec:mockobsresult}

\begin{figure*}[htbp]
    \centering
    \includegraphics[width=7in]{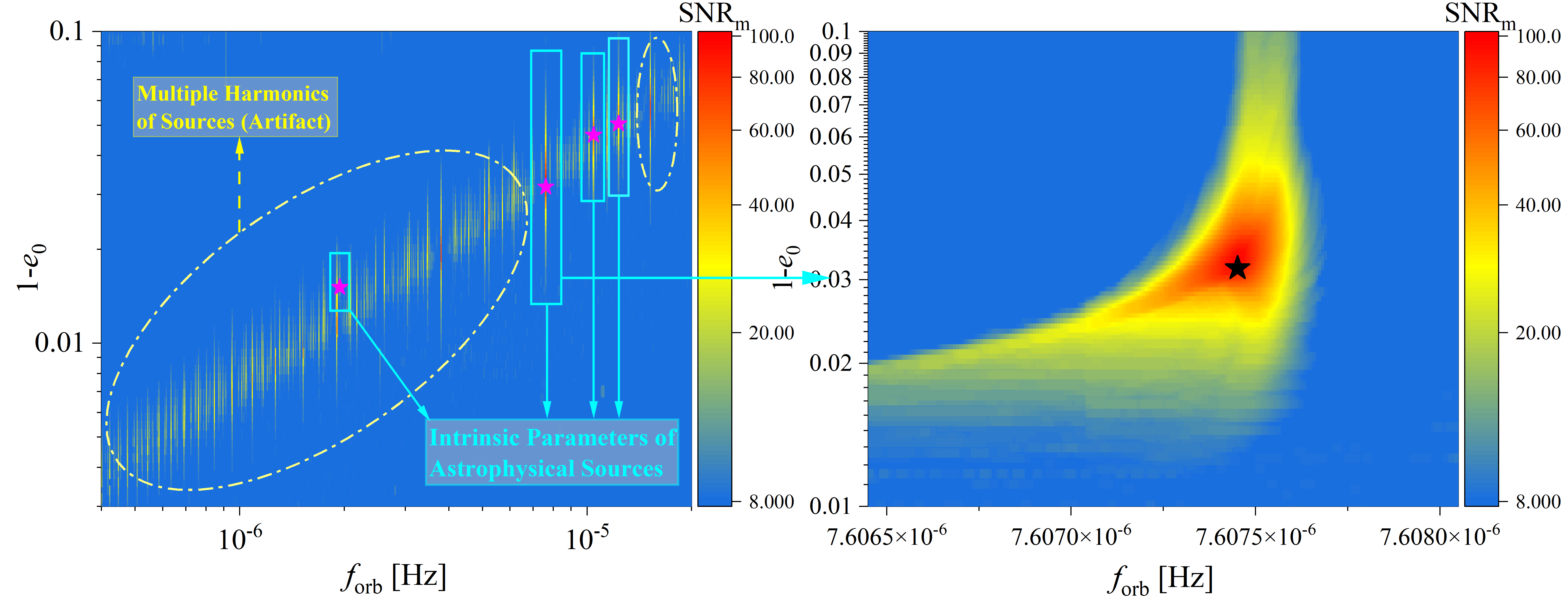} 
    \caption{{\bf The matched filtering output in the mock LISA data analysis of highly eccentric BBHs.} Here we fit the mock LISA data with different bursting GW templates and plot the optimal signal-to-noise ratio, $\rm SNR_m$, in different colors, as a function of $f_{\rm orb}$ and $1-e_0$ of the template (see Sections~\ref{sec:waveform_model} and \ref{sec:mockobservation} for more details). Specifically, the optimal $\rm SNR$ is calculated numerically, using matched filtering as described in Section~\ref{sec:mockobservation} (note that $\rm SNR_m$ represents the realistic fitting result, and is different from the analytical estimation of $\rm SNR$ in Equation~(\ref{eq:SNR}) and Figures~\ref{fig:fishernum} - \ref{fig:population}). {\it Left Panel} shows the fitting results for all the potential GW sources with an orbital frequency of $f_{\rm orb}\sim 4\times 10^{-7}-2\times 10^{-5}$~Hz and eccentricity $e\sim 0.9-0.997$, while the {\it Right Panel} shows a zoom-in of the region near the parameters of a ``real" GW source that appears in the mock data (i.e., the detection case). For demonstration purposes, we mark, in stars, the intrinsic parameters of the GW sources and highlight the region where we successfully fit these signals with high $\rm SNR_m$ (see the rectangular boxes in light blue). We note that, in some regions of {\it Left Panel}, numerical artifacts cause the fitted $\rm SNR_m$ to exceed the detection threshold of $8$, while there are no astrophysical sources (see Section~\ref{sec:mockobservation} for a more detailed discussion).
    \footnote{We applied a high-pass filter in the data analysis before the matched filtering to suppress the loud instrumental noise and astrophysical foreground at frequencies below $10^{-4}$ Hz. Additionally, to reduce computational costs, we generated mock LISA data with a sample rate of 0.1 Hz. Note that the matched filtering produces a $\rm SNR$ time series for each template (see, e.g., Equation~(\ref{eq:MF})). To overcome numerical artifacts caused by the finite length of the GW signal, we implemented a cutoff in the SNR time series at $t_0< 5/f_{\rm orb}$ and $t_{\rm end}-t_0< 5/f_{\rm orb}$ (in which $t_{\rm end}=1$~yr is the ending time of the GW signal), before finding the $\rm SNR_m$ and outputting the result. A similar approach can be found in LIGO data analysis, see, e.g., Sec.VII in Ref.~\citep{Allen2011}. }}
    \label{fig:mfresult}
\end{figure*}

In the mock data analysis, we adopt the matched filtering method described in Section~\ref{sec:MFmethod}. Specifically, we first generate a waveform template bank using Equations~(\ref{eq:strains}) - (\ref{eq:response2}), which include the effect of the detector's response and annual motion around the Sun. For simplicity, we fix the mass of the template binaries as $10 - 10~{\rm M_{\odot}}$, with $\theta=\phi=\Theta=\Phi=\psi=\pi/4$, and change the initial orbital frequency $f_{\rm orb}$ and eccentricity $e_0$. We then convolve each template signal with the mock LISA data and compute the optimal signal-to-noise ratio, $\rm SNR_m$ (see Section~\ref{sec:MFmethod}, and Eq.~(\ref{eq:SNRm})). We remind the reader that the PyCBC software package is adopted for the numerical realization \citep{PYCBC_alex_nitz_2024_10473621}. The results of the Match filtering output are shown in Figure~\ref{fig:mfresult}.

We emphasize that, as discussed above, some parameters of the waveform template are fixed to avoid high computational costs in the mock data analysis. Specifically, we only scan the 2-dimensional parameter space of $(f_{\rm orb}, 1-e_0)$, while the template signals may not match the exact values of $(M,q,\cos\Theta,\Phi,\cos\theta,\phi,\psi)$ of the GW sources. (Note that matched filtering is independent of the overall amplitude of the waveform, so the distance $R$ of the template binary will not affect the fitting result, assuming redshift is negligible.) Consequently, the $\rm SNR_m$ we show in Figure~\ref{fig:mfresult} does not represent a global optimum for the entire 10-dimensional parameter space (see Section~\ref{sec:waveform_model}). However, we expect such a loss of signal-to-noise ratio will not significantly impact the overall fitting result, as the bursting waveform is less sensitive to $(M,q,\cos\Theta,\Phi,\cos\theta,\phi,\psi)$ compared to $(f_{\rm orb}, 1-e_0)$ (see, e.g., the Fisher matrix analysis result in Section~\ref{subsec:fishernum}; note that the parameters with larger measurement error will have a weaker influence on the waveform). For example, the GW source with $a\sim 0.068$~au, $(1-e)\sim 0.032$ has an analytical estimation of $\rm SNR\sim 200$ in Figure~\ref{fig:population} (see the biggest black dot). On the other hand, this source is successfully identified in the mock data analysis, numerically fitted with $\rm SNR_m=96$ and almost no bias in the measured values of $(f_{\rm orb}, 1-e_0)$ (see the black star in the {\it Right Panel} of Figure~\ref{fig:mfresult}). Therefore, even without fitting the global optimum across the 10-dimensional parameter space, we can effectively apply the matched filtering to search for $(f_{\rm orb},1-e_0)$, which are the dominant parameters that affect the shape of bursting waveform, and still detect most of the bursting sources with high $\rm SNR$ in future LISA observation.

In Figure~\ref{fig:mfresult}, we plot the fitted $\rm SNR_m$ in different colors as a function of the template's $f_{\rm orb}$ and $1-e_0$. As shown in the {\it Left Panel}, we successfully identify multiple bursting sources in the mock data. In particular, the stars represent the intrinsic parameters of four GW sources that have a high analytical $\rm SNR$ in the mock data (note that these sources, as well as other low $\rm SNR$ sources in Figure~\ref{fig:population}, are present simultaneously in the mock LISA signal). After applying the matched filtering, we get the $\rm SNR_m$ distribution as highlighted by the rectangular boxes in light blue, which has a clear high $\rm SNR_m$ peak that centers around the intrinsic parameters of the binary (with the peak $\rm SNR_m=15,96,54,47$, from left to right, exceeding the detection threshold of 8). This result verifies that matched filtering can perform well in extracting the ``bursting" GW signal from LISA data, even when multiple sources are contributing to a wide range of harmonics in the frequency domain (see Section~\ref{sunsec:differenceofburstfitting}) and when there are instrumental noises plus the Galactic foreground.

Furthermore, {\it Right Panel} illustrates a zoom-in of the region near the intrinsic parameters of a detected source. We emphasize that the matched filtering of bursting GW signal is highly sensitive to the choice of the initial orbital frequency, $f_{\rm orb}$. For example, the high $\rm SNR_m$ peak in {\it Left Panel} only spans a frequency range of $\sim 10^{-10}$~Hz. This sensitivity implies that, for LISA data analysis, we need to choose a grid size smaller than this value (or, more precisely, as estimated in Equation~(\ref{eq:analyticalestiforb})) to avoid missing any bursting sources in the search. Given the parameter space scale of $f_{\rm orb}\sim 10^{-7}-10^{-4}$~Hz, this grid size requires computing $\sim 10^6$ grids in the dimension of $f_{\rm orb}$, which greatly increases the computational expense. However, once the sources are identified, we can take advantage of this high sensitivity to $f_{\rm orb}$ and accurately determine the orbital parameters of bursting binaries. Moreover, because of the small error in orbital frequency measurement, two bursting sources with similar orbital frequencies are less likely to be confused in the data analysis, which could help suppress any potential confusion background.



We note that there are numerical artifacts in the {\it Left Panel} of Figure~\ref{fig:mfresult}, which cause the fitted $\rm SNR$ to exceed the detection threshold of 8 even when there are no astrophysical sources. Specifically, in the region surrounded by yellow dash-dotted lines, most of the high $\rm SNR_m$ peaks do not represent real parameters of astrophysical sources (except for the four $\rm SNR_m$ peaks highlighted by rectangular boxes in light blue). Instead, these peaks are caused by the higher-order harmonics of astrophysical bursting signals. Specifically, for a given astrophysical signal with burst period $1/f_{\rm orb}$, all the templates $h_{\rm template}(t)$ with the following orbital frequencies could have high $\rm SNR_m$ in the template fitting:
\begin{equation}
f_{\rm orb}^{\prime} =
     \frac{m}{n} f_{\rm orb}, \, 
     (m,n=1,2,3...) 
\label{eq:fakeharmonics}   
\end{equation}
Equation~(\ref{eq:fakeharmonics}) can be explained using the property of bursting signals. In particular, most of the observed stellar mass bursting systems are expected to have slow orbital evolution (see Section~\ref{sec:astro}), and their GW emission can be characterized by repeated bursts, with a separation of $\delta\tau=1/f_{\rm orb}$ in the time domain. For example, when using a template with a burst separation of $2\delta\tau,\,3\delta\tau,\, 4\delta\tau...$ or $\frac{1}{2}\delta\tau,\,\frac{1}{3}\delta\tau,\, \frac{1}{4}\delta\tau...$ to fit the signal, the GW bursts of the template will also match (some of) the bursts from astrophysical sources, thus yielding a high $\rm SNR_m$ in the matched filtering result (note that, in the frequency domain, this effect can be described as the overlapping of harmonics between two signals with different fundamental frequencies, with one fundamental frequency being multiple integer times the other one; \xzy{Also, a similar effect has been found in the literature of EMRI data analysis, see, e.g., \citet{Chua_2022}}). 

Additionally, to excite the highest $\rm SNR_m$ peak at a given harmonic frequency $f^{\prime}_{\rm orb}$, the templates need to have most of the GW energy distributed in the same frequency band as the astrophysical source. In other words, the template should have a burst frequency that closely matches the astrophysical GW signal (see Equation~(\ref{eq:fpeak})):
\begin{equation}
    f^{\prime}_{\rm orb}(1-e^{\prime})^{-\frac{3}{2}}\sim f_{\rm orb}(1-e_0)^{-\frac{3}{2}}\, ,
    \label{eq:fburstfit}
\end{equation}
in which $e^{\prime}$ is the eccentricity of the harmonic template. Notably, Equation~(\ref{eq:fburstfit}) also suggests that the width of a GW burst in the template coincides with the width of the burst in the actual astrophysical signal.

Equations~(\ref{eq:fakeharmonics}) and (\ref{eq:fburstfit}) provide a prediction for the orbital frequency and eccentricity of the harmonic template (see, e.g., the harmonic peak locations in Figure~\ref{fig:mfresult}), which can significantly enhance source identification in future data analysis. For example, the system with $f_{\rm orb}=7.60745\times 10^{-6}$~Hz will excite multiply harmonics at the frequencies such as $3.80372, 2.53581,1.90186\times 10^{-6}$~Hz and $1.52149, 2.28223,3.04298\times 10^{-5}$~Hz (which represent the $2$nd, $3$rd, and $4$th harmonics, see the peaks in the {\it Left Panel} of Figure~\ref{fig:mfresult}). Once finding these high $\rm SNR_m$ peaks, we can try to fit their locations using Equations~(\ref{eq:fakeharmonics}) and (\ref{eq:fburstfit}), thus extracting the intrinsic parameters of the source. We note that, the fitted $\rm SNR_m$ will be suppressed as the harmonic numbers $m,n$ increase, resulting in a finite number of detectable harmonics. These harmonics can be efficiently distinguished due to the high accuracy of $f_{\rm orb}$ measurement. Furthermore, Figure~\ref{fig:mfresult} shows the matched filtering result for simultaneously fitting all bursting sources. However, in practical data analysis, we can identify the loudest bursting sources, remove them from the dataset, and thereby prevent their harmonics from interfering with the fitting of other astrophysical signals.

Here, we aim at a proof-of-concept example of a bursting source search in LISA data analysis. Thus, this mock observation does not include all the potential bursting source populations. Furthermore, we assume a stationary Gaussian instrumental noise in the mock LISA data, which may not capture all the influence of noise on the fitting results of the bursting signals. For example, there have been many studies highlighting the importance of instrumental glitches \citep[see, e.g., ][]{Robson_2019glitches,LISApathfinder22glitch,LISApathfinder22transientacc}. We expect this phenomenon to appear in future detection, as shown recently by the LISA Pathfinder (see, e.g., Refs.~\citep{LISApathfinder16,LISApathfinder16b}). Therefore, it is important to characterize whether these noise transients will affect LISA's performance in detecting highly eccentric compact binaries, which we leave for future works. 

Additionally, as discussed above, the matched filtering of bursting GW signals is very sensitive to the orbital frequency, with $\sim 10^6$ grids required for a complete search of $f_{\rm orb}$ in the parameter space that we focus on. Moreover, there are 9 more dimensions $(1-e_0, M, q, \cos\Theta,\Phi,\cos\theta,\phi, R,\psi)$ that contribute to the total number of grids we need to compute. However, computing bursting waveforms in the time domain is time-consuming, which leads to a high computational cost for the search of these sources. Thus, we highlight the importance of constructing accurate frequency-domain waveforms for highly eccentric binaries, as they could greatly accelerate realistic data analysis. Also, we may need to incorporate other sub-optimal time-frequency approaches to search for the signal, such as power stacking \citep{east13}, the TFCLUSTER algorithm \citep{sylvestre02}, wavelet decomposition \citep{Klimenko_2004}, and the Q-transform \citep{Chatterji2004Qtransform,Chatterji2006searchburst,bassetti2005development,Tai_2014}, which are more robust to different kinds of transients and have lower computational costs.

\section{Discussion}
\label{sec:discussion}

Wide, highly eccentric compact binaries can naturally arise as a progenitor stage of GW mergers, particularly in dynamical channels where environmental perturbations bring two compact objects into close encounters \citep[see, e.g.,][]{O'Leary+09,Thompson+11,Aarseth+12,Kocsis_2012,breivik16,Gondan_2018a,Orazio+18,Zevin_2019,Samsing+19,Martinez+20,Antonini+19,Kremer_2020,wintergranic2023binary,Gondan_Kocsis2021,Michaely+19,Michaely+20,Michaely+22,wen03,Hoang+18,Hamers+18,Stephan+19,Zevin_2019,Bub+20,Deme+20,Wang+21,Zevin_2021,Xuan+21,Xuan+23b,Mockler+23SMBHtde,Melchor+24TDEcombinedeffect,Xuan24bkg}. These systems will undergo effective GW emission upon each pericenter passage, which potentially creates a burst-like pattern in the mHz GW detection, as demonstrated in Figure~\ref{fig:egwaveform} (note that there could be $\sim 3 - 45$ detectable bursting BBHs in the Milky Way, which contribute to $\sim 10^2 - 10^4$ GW bursts during the LISA mission \citep{Xuan+23b}). This work focuses on the source identification and parameter extraction of stellar mass bursting binaries. Particularly, we explored the astrophysical inference of these highly eccentric systems, quantified their parameter measurement error, and carried out mock LISA observation to test whether these bursting binaries could be efficiently distinguished when multiple sources and instrumental noises are present simultaneously in the detector’s output.

Throughout this paper, we have adopted the x-model to generate GW signals for all of the bursting binaries (see Section~\ref{sec:astro}). This model includes the dynamics of eccentric compact binaries to $3$~pN order, and we further account for the detector's annual motion around the Sun (see Equations~(\ref{eq:strains}) - (\ref{eq:response2})). The definition of coordinates is illustrated in Figure~\ref{fig:coordinate}.

To assess the detectability and parameter measurement accuracy of bursting GW sources in the LISA band, we utilize the Fisher matrix analysis (see Section~\ref{sec:fisherana}) and the matched filtering method (see Section ~\ref{sec:MFmethod}). 
In particular, using the Fisher matrix analysis, we compute the relative measurement error of $\left\{f_{\rm orb}, 1-e_0, M, q, \cos\Theta, \Phi, \cos\theta,\phi, R,\psi \right\}$, for the bursting GW signals from a $10-15\,{\rm M_{\odot}}$ BBH system at a distance of $8$~kpc, in a representative parameter space of semi-major axis $a_0\sim 0.01-3$~au, eccentricity $e_0\sim 0.9-0.9999$. The results are shown in Figures~\ref{fig:fishernum} and \ref{fig:fishernum3}, with Figure~\ref{fig:fishernum} representing 1-year LISA observation and Figure~\ref{fig:fishernum3} representing 5 years (which yields a higher accuracy).

As illustrated in Figures~\ref{fig:fishernum} and \ref{fig:fishernum3}, for a highly eccentric binary system, its orbital frequency, eccentricity (as $1-e_0$), and the sky location can be retrieved with relatively high accuracy. Notably, LISA could determine these bursting sources' orbital frequency to an accuracy of $\Delta f_{\rm orb}\sim 10^{-12}-10^{-10}$~Hz, with the relative error of $1-e_0$ measurement smaller than $\sim 1\%$ in the detected cases, which could greatly help with the understanding of these binaries' formation channels. Furthermore, the sky location of bursting sources could be measured to an accuracy of $\sim 0.1-10$~degrees (depending on the signal-to-noise ratio), while the bursting waveform detection may be less sensitive to other parameters, including the binary's intrinsic orientation $\{\Theta, \Phi\}$, the distance $R$, and the polarization angle $\psi$ of GW signal. Additionally, the mass ratio $q$ is poorly constrained, but we could estimate the total mass $M$ with an accuracy of $\Delta M \lesssim 10\%$ in a $5$~yr observation, when the $\rm SNR$ exceeds $\sim 100$. We discussed the astrophysical interpretation of the aforementioned results in Section~\ref{subsec:astrointerpretation}, and give an analytical estimation of $f_{\rm orb}$ measurement error in Equation~(\ref{eq:analyticalestiforb}).

\xzy{We note that, due to the limitation of the Fisher matrix analysis, the aforementioned parameter measurement accuracy only serves as a heuristic estimation for this new source class. In particular, the resultant measurement error of our work can become unreliable when compared with full Bayesian analyses (see, e.g., the discussion in Ref.~\citep{Vallisneri_2008}), especially for those parameters with weak influence on the waveform (i.e., mass ratio $q$, intrinsic orientation $\Theta,\,\Phi$, and distance $R$). Thus, we emphasize the need for full Bayesian analyses in future work. While computationally expensive, these methods can consistently include sophisticated priors, and explore the secondary maxima of the posterior. Nonetheless, our results provide compelling evidence that bursting GW sources are promising candidates for inferring astrophysical properties.}

Moreover, to measure the source parameters in realistic observation, the bursting GW signals need to be properly identified and extracted from the data stream. Thus, we further test the detectability of the highly eccentric compact binaries via the mock LISA data analysis. Specifically, we construct an example of mock LISA data (see Section~\ref{subsec: population}), which includes a representative population of bursting BBHs in the Milky Way Galactic center nuclear star cluster (see Figure~\ref{fig:population}), the instrumental noise of detector, and the astrophysical GW noise foreground
caused by unresolved Galactic binaries. We then adopt the matched filtering method as described in Section~\ref{sec:MFmethod}, to compute the optimal signal-to-noise ratio in the search of bursting GW sources.



The mock data analysis results are summarized in Figure~\ref{fig:mfresult}. Interestingly, even without fitting the global optimum in the 10-dimensional parameter space, the matched filtering search of bursting GW templates in the plane of $(f_{\rm orb},1-e_0)$ still identifies most
of the bursting sources with high SNR in the mock observation (which is mostly because $(f_{\rm orb},1-e_0)$ are the dominant parameters that characterize the bursting GW signals). Furthermore, the matched filtering results show a high sensitivity to orbital frequency (see the {\it Left Panel}), which indicates a small grid size requirement, $\Delta f_{\rm orb}\sim 10^{-10}$~Hz, for future data analysis. On the other hand, this phenomenon also means there could be less confusion between bursting binaries with different values of $f_{\rm orb}$. Thus, multiple sources can be distinguished simultaneously in the detector's output. We also explore the potential artifacts in burst data analysis, such as the non-astrophysical signals caused by the higher-order harmonics of highly eccentric GW sources, as described in Equation~(\ref{eq:fakeharmonics}).

We emphasize that the mock data analysis in this work does not include all the potential GW source populations and the non-Gaussian instrumental noise (such as glitches). However, it serves as a proof-of-concept example of bursting source search in LISA data analysis, which verifies the detectability of highly eccentric compact binaries with the current LISA design and data analysis methods. Also, our analysis provides a guideline for the future bursting GW signal search, including an estimation of parameter space, computational expense, and potential artifacts.  

To conclude, wide, highly eccentric compact binaries can provide us with valuable information about their formation channels and the surrounding environment. In particular, we show that these sources could be properly identified and extracted from the LISA detector's output, with high accuracy in the measurement of orbital frequency, eccentricity, and sky location. However, the computational expense may be large for the search of bursting signals. Therefore, it is important to develop ready-to-use frequency domain templates for bursting GW sources or incorporate other efficient time-frequency approaches in future detection.

\acknowledgments
The authors thank Aditya Vijaykumar for the valuable discussions and the anonymous referee for their useful comments. ZX acknowledges partial support from the Bhaumik Institute for Theoretical Physics summer fellowship. ZX and SN acknowledge the partial support from NASA ATP 80NSSC20K0505 and from NSF-AST 2206428 grant and thank Howard and Astrid Preston for their generous support. Further, ZX and SN thank LISA Sprint 2024 for organizing an interactive and productive meeting. JM acknowledges support from the Canada Research Chairs program. AMK acknowledges funding support from a Killam Doctoral Scholarship. This work was supported by the Science and Technology Facilities Council Grant Number ST/W000903/1 (to BK).

\bibliographystyle{apsrev4-1.bst}
\bibliography{bibbase}

\appendix
\section{Time Domain Estimation of the Inner Product}
\label{app:timedomain}
For quasi-periodic gravitational waves $h_1,h_2$ with frequency close to $f_0$, the GW signal is a nearly monochromatic, sinusoidal waveform. In this case, the frequency-domain expression of inner product (see Equation~(\ref{eq:innerproduct})) can be simplified. In particular, the noise curve $S_n(f)$ can be taken out of the integration because of the small variation of $f$:
\begin{equation}
\left\langle h_1 \mid h_2\right\rangle \approx \frac{2}{S_n(f_0)} \int_0^{\infty}\left[\tilde{h}_1^*(f) \tilde{h}_2(f)+\tilde{h}_1(f) \tilde{h}_2^*(f)\right] d f . \label{eq:innerproduct2}
\end{equation}
Furthermore, using Parseval's theorem, we can compute the integral in Equation~(\ref{eq:innerproduct2}) using time domain waveforms:
\begin{equation}
\left\langle h_1 \mid h_2\right\rangle \approx \frac{4}{S_n(f_0)} \int_0^{\infty} h_1(t) h_2(t) d t .\label{eq:timeinnerproduct}
\end{equation}
Equation~(\ref{eq:timeinnerproduct}) is commonly adopted when analyzing the gravitational waves from circular binaries when the evolutionary timescale is much longer than the observational period of LISA (see, e.g., \citep{chen19,seto02}\footnote{Note that Eq.~ 22 in \citet{seto02} differs from the value of frequency inner product in Eq.~2.3 of \citet{Cutler+94} by a factor of 2, to recover the value of signal-to-noise ratio the consistent coefficient in the time domain inner product should be 4 instead of 2.}).

\section{Analytical Estimation of the Fisher Matrix}
\label{appendix:fisher}

In this section, we develop an analytical method to estimate the Fisher matrix for eccentric GW sources in the LISA band. In particular, we first adopt a simplified waveform model for eccentric sources from Ref.~\citep{Moreno+95} (see also \citep{Seto+01,Mikoczi+12,Gondan_2018a}), in which $h_{\times},\,h_+ $ denote two polarizations of the gravitational wave and $n=1,2,3...$ denotes the number of harmonics:

\begin{align}
h_{\times}(t)= & -h \cos \Theta \sum_n\left[B_{n-} \sin \Phi_{n+}^t+B_{n+} \sin \Phi_{n-}^t\right], \label{eq:eccwaveform0}\\
h_{+}(t)= & -\frac{h}{2} \sum_n\left[\sin ^2 \Theta A_n \cos \Phi_n^t+\left(1+\cos ^2 \Theta\right)\right.  \nonumber\\
& \left.\times\left(B_{n+} \cos \Phi_{n-}^t-B_{n-} \cos \Phi_{n+}^t\right)\right] .\label{eq:eccwaveform}
\end{align}
here $h=4 m_1m_2/\left(a R\right)$ is the strain amplitude of gravitational wave. For a detailed description of orbital parameters, see Fig.1 in Ref.~\citep{Mikoczi+12}.

In Equation~(\ref{eq:eccwaveform}), $B_{n \pm}=$ $\left(S_n \pm C_n\right) / 2$ and $A_n$ are linear combinations of the Bessel functions of the first kind, $J_n(n e)$, and their derivatives:
\begin{align}
S_n&=-\frac{2\left(1-e^2\right)^{1 / 2}}{e} {n^{-1}} J_n^{\prime}(n e)+\frac{2\left(1-e^2\right)^{3 / 2}}{e^2} n J_n(n e), \\\nonumber
C_n&=-\frac{2-e^2}{e^2} J_n(n e)+\frac{2\left(1-e^2\right)}{e} J_n^{\prime}(n e), \\\nonumber
A_n&=J_n(n e),
\end{align}
where a prime denotes the derivative, i.e., $J_n^{\prime}(n e) \equiv dJ_n(n e)/de=n\left[J_{n-1}(n e)-J_{n+1}(n e)\right] / 2$. 

The frequency component of an eccentric GW waveform is made up of a series of harmonics, with the power of each harmonic concentrate near $f_n=nf_{\rm orb}$ \citep[see, e.g., ][]{Peters64,Seto+01,Xuan23acc,Xuan+23b,Xuan24bkg}. Furthermore, the pericenter precession makes the orbital phase of each harmonic split into a triplet, $\Phi_n,\, \Phi_{n\pm}$. Therefore, in Equation~(\ref{eq:eccwaveform}), the time evolution of orbital phases can be expressed as \citep{Seto+01}: 
\begin{equation}
\Phi_n^t=2 \pi n f_{\text {orb }} (t-t_0), 
\Phi_{n \pm}^t=2 \pi n f_{\text {orb }}  (t-t_0) \pm 2 \pi \delta f t \pm 2 \gamma_0,
\label{eq:Phi}
\end{equation}
in which $f_{\rm orb}=T_{\rm orb}^{-1}$ is the Keplerian orbital frequency (here $T_{\rm orb}=$ $2 \pi M^{-1 / 2} a^{3 / 2}$ is the Newtonian radial orbital period), $t_0$ is the time of pericenter passage (hereafter, we set $t_0=0$),  $\gamma_0$ is the initial angle of the pericenter, and here we define $\delta f$ as twice of the (general relativistic) pericenter precession frequency \cite{Seto+01}:
\begin{equation}
\delta f= 2\times \frac{1}{2\pi} \frac{d\gamma}{dt}=\frac{3 M^{\frac{3}{2}}}{\pi a^{\frac{5}{2}}\left(1-e^2\right)}=\frac{6 (2 \pi )^{\frac{2}{3}}M^{\frac{2}{3}}f_{\rm orb}^{\frac{5}{3}}}{\left(1-e^2\right)} .  
\end{equation}

After setting up the waveform, we can take its derivative of different orbital parameters and get an analytical estimation of the Fisher matrix using Equation~(\ref{eq:fisherdefinition}). In the most general case, all the six parameters could affect the waveform in Equation~(\ref{eq:eccwaveform}), i.e., $(h, \Theta, f_{\rm orb}, \delta f, e, \gamma_0)$. However, for simplicity, here we will only focus on three parameters, i.e., $(f_{\rm orb}, \delta f,e)$. This simplification is partly justified because $(f_{\rm orb}, \delta f,e)$ dominates the long-term phase evolution of the GW waveform, thus mostly affecting the overlap in the computation of Fisher matrix (see Equation~(\ref{eq:fisherdefinition})). Also, we are interested in these three parameters because they represent the accuracy of measurement for the (redshifted) intrinsic parameters of the binary system.

\begin{align}\label{eq:partialforb}
\frac{\partial h_{\times}}{\partial f_{\text {orb }}}=-h \cos \Theta \sum_n & \left[B_{n-}\cos \Phi_{n+}^t \cdot(2 \pi n t)\right. \\\nonumber
& \left.+B_{n+} \cos \Phi_{{n-}}^t \cdot(2 \pi n t)\right]
\end{align}
\begin{align}\label{eq:partialdeltaf}
 \frac{\partial h_\times}{\partial \delta f}=-h \cos \Theta \sum_n & \left[B_{n-} \cos \Phi_{n+}^t \cdot(2 \pi t)\right. \\\nonumber
& \left.+B_{n+} \cos \Phi_{{n-}}^t \cdot(-2 \pi t)\right] 
\end{align}
\begin{align}
\frac{\partial h_{\times}}{\partial e}=-h \cos \Theta \sum_n\left[B_{n-}^{\prime} \sin \Phi_{n t}^t+B_{n+}^{\prime} \sin \Phi_{{n-}}^t\right]
\end{align}
in which $B_{n\pm}^{\prime}(n,e)=dB_{n\pm}/de$ is the derivative as a function of the binary's eccentricity. Note that, in Equations~(\ref{eq:partialforb}) and (\ref{eq:partialdeltaf}), the factor of $2\pi n t$ ($ 2\pi t$) is because of taking the derivative of $\Phi_{n\pm}^t$ over $f_{\rm orb}$ ($\delta f$) (see Equation~(\ref{eq:Phi})).

We note that, on the other hand, $h$ represents the overall amplitude of the waveform, while $\Theta$ and $\gamma_0$ describe the inclination and orientation of the binary relative to the observer. Since there are degeneracies between some of these parameters (e.g., $h$ and $\cos \Theta$ in Equation~(\ref{eq:eccwaveform0})), and the highly eccentric GW waveform is less sensitive to these parameters (see e.g., they are weakly constrained in Figure~\ref{fig:fishernum}), we neglect their contribution for simplicity. However, in realistic observation, these parameters may still interfere with the measurement of $(f_{\rm orb}, \delta f, e)$. In other words, our result in the following discussion only considers the submatrix of $(f_{\rm orb}, \delta f, e)$, and the corresponding estimation serves as a lower bound of parameter measurement error.

Here, we assume the GW source evolves slowly. Thus, the signal does not undergo a significant frequency shift during the observation, and each harmonic is a near-monochromatic signal with the power concentrates near $f_n=nf_{\rm orb}$. We expect that most sources will be consistent with this assumption for the harmonics that contribute most to the detection signal-to-noise ratio \citep[e.g.,][]{O'Leary+09,Wang+21,Xuan+23b}; however, in some cases, the signal may shift in a short timescale, or the precession of orbit is fast \citep{Hoang+19,Deme+20}, which is beyond the scope of this study.  

In particular, we make the following assumptions:
\begin{align}
\left\{
\begin{array}{cc}
     \frac{f_{\rm orb}}{\dot{f}_{\rm orb}}\gg T_{\rm obs} &\\[3ex]
     T_{\rm obs}\gg \frac{1}{f_{\rm orb}} &\\[3ex]
     f_{\rm orb}\gg \delta f &
\end{array}
\right.
\label{eq:approximation}
\end{align}
here $\dot f_{\rm orb}$ represents the time derivative of orbital frequency, which can be estimated using \citep{Peters64,Xuan+21}:
\begin{equation}
	\dot{f}_{\rm orb}=\frac{1}{2}\times
    \frac{96 \pi^{8/3}}{5}\mathcal{M}_{c}^{5/3}\left(2 f_{\rm orb}\right)^{11/3}F(e) \ ,
    \label{eq:chirp mass}
\end{equation}
in which $\mathcal{M}_{c}$ is the chirp mass of the binary, ${\cal
M}_{c}\equiv (m_1m_2)^{3/5}/(m_1+m_2)^{1/5}$, $F(e)$ is a function of the compact binary's eccentricity (the enhancement function)\citep{peters63}:
\begin{equation}\label{eq:Fe}
F(e)\equiv\frac{1+\frac{73}{24}e^{2}+\frac{37}{96}e^{4}}{(1-e^{2})^{7/2}} \ .
\end{equation}

In Equation~(\ref{eq:approximation}), the first assumption ensures that the source evolves slowly (merger timescale $\tau\sim f_{\rm orb}/\dot{f}_{\rm orb}\gg T_{\rm obs}$), the second assumption ensures that there is a significant number of bursts detected during the observation, and the third assumption ensures that pericenter precession is small during each orbit (i.e., every single harmonic and its corresponding triplet frequencies will only overlap with itself in the frequency domain, see Equation~(\ref{eq:Phi})).
 
Therefore, under the assumption that $f_{\rm orb}$ evolves slowly and $\delta f$ is much smaller than $f_{\rm orb}$, each single harmonic (and its corresponding triplet frequencies) can be taken as a near-monochromatic signal, which does not overlap with other harmonics. This property allows us to compute the inner product of each individual harmonic using Equation~(\ref{eq:timeinnerproduct}), then sum over all the harmonics' contributions to get the total inner product.

For example, combining Equations~(\ref{eq:timeinnerproduct}) and (\ref{eq:partialforb}), we have:
\begin{align}
 \left\langle\left.\frac{\partial h_\times}{\partial f_{\text {orb }}} \right\rvert\, \frac{\partial h_\times}{\partial f_{\text {orb }}}\right\rangle & =h^2 \cos ^2 \Theta \cdot \sum_n \frac{4}{S_n(nf_{\text {orb }})} \cdot(2 \pi n)^2 \nonumber\\ 
& \times\int_0^{T_{\rm obs}}\left[B_{n-}^{2} \cos ^2 \Phi_{{n+}}^t+B_{n+}^{2} \cos \Phi_{{n-}}^2\right. \nonumber\\
& \left.+2 B_{n+} B_{n-}\cos \Phi_{{n+}}^t \cos \Phi_{n-}^t\right] t^2 d t \ ,\label{eq:time_inner_partial}
\end{align}
in which:
\begin{align}
& \int_0^{T_{\rm obs}} \cos ^2 \Phi_{n\pm}^t t^2 d t =\frac{1}{2}\int_0^{T_{\text {obs }}} (1+\cos  2\Phi_{n\pm}^t) t^2 d t \nonumber\\
& =\frac{1}{6} T_{\rm obs}^3+\frac{1}{2}\int_0^{t_{\text {obs }}} \cos \left[4 \pi\left(n f_{\rm orb} \pm \delta f\right) t \pm 4 \gamma_0\right] t^2 d t \nonumber\\
& \simeq \frac{1}{6} T_{\rm obs}^3(1+\mathcal{O}(1))\ ,\label{eq:selfproduct}
\end{align}
and:
\begin{align}
& \int_0^{t_{\rm obs }} \cos \Phi_{n+}^t \cos \Phi_{{n-}}^t t^2 d t \label{eq:crossterm}\\
&=\frac{1}{2} \int_0^{T_{\rm obs}}\left[\cos \left(4 \pi n f_{\rm orb} t\right)+\cos \left(4 \pi \delta f t+4 \gamma_0\right)\right] t^2 dt\ . \nonumber
\end{align}

We note that, the integration of $\cos \left(4 \pi \delta f t+4 \gamma_0\right) t^2 dt$ in Equation~(\ref{eq:crossterm}) may yield a significant contribution to the inner product. This is because many highly eccentric systems have their precession timescale $1/\delta f$ comparable to or longer than the observation time, $T_{\rm obs}$. Therefore, Equation~(\ref{eq:crossterm}) can be on the same order of $\sim T_{\rm obs}^3$. (Note that we didn't assume $T_{\rm obs}\gg 1/\delta f$ in Equation~(\ref{eq:approximation}), otherwise, all the terms in Equation~(\ref{eq:crossterm}) can be neglected compared with $T_{\rm obs}^3$.)

In other words, the initial angle of the pericenter, $\gamma_0$, can affect the parameter extraction, providing that the eccentric GW source has slow precession during the observation. However, since here we are focusing on a heuristic estimation of the parameter extraction accuracy, we further average over different orientations of the source, i.e., the angle of $\gamma_0$, to get the general order of magnitude for the matrix element. The average over $\gamma_0$ makes the term of $\cos \left(4 \pi \delta f t+4 \gamma_0\right) t^2 dt$ vanish. Therefore, after taking this average, the terms in Equation~(\ref{eq:crossterm}) can be neglected because they are much smaller than $T_{\rm obs}^3$ (the leading order contribution, see e.g., Equation~(\ref{eq:selfproduct})). 

Plug the result of Equations~(\ref{eq:selfproduct}) and (\ref{eq:crossterm}) into Equation~(\ref{eq:time_inner_partial}), we can get:
\begin{align}\label{eq:gammaforbcorss}
&\left\langle\Gamma^{\times}_{{f_{\rm orb}f_{\rm orb}}} \right\rangle_{\gamma_0} =\left\langle\left.\frac{\partial h_\times}{\partial f_{\text {orb }}} \right\rvert\, \frac{\partial h_\times}{\partial f_{\text {orb }}}\right\rangle_{\gamma_0}  \nonumber\\ &\sim \frac{8\pi^2}{3} h^2 \cos ^2 \Theta  T_{\rm obs}^3  \sum_n \frac{n^2}{S_n(nf_{\text {orb }})} ( B_{n+}^2 + B_{n-}^2) 
\end{align}
in which $\langle\cdot\rangle_{\gamma_0}$ represents taking the average over $\gamma_0$.

Similarly, we can adopt this method for the other two parameters, $\delta f$ and $e$, and get: 
\begin{align}
&\left\langle\Gamma^{\times}_{\delta f \delta f} \right\rangle_{\gamma_0} =\left\langle\left.\frac{\partial h_\times}{\partial \delta f} \right\rvert\, \frac{\partial h_\times}{\partial \delta f}\right\rangle_{\gamma_0} \nonumber\\ &\sim \frac{8\pi^2}{3} h^2 \cos ^2 \Theta  T_{\rm obs}^3  \sum_n \frac{1}{S_n(nf_{\text {orb }})} ( B_{n+}^2 + B_{n-}^2) 
\end{align}
\begin{align}
&\left\langle\Gamma^{\times}_{ee}\right\rangle_{\gamma_0} =  \left\langle\left.\frac{\partial h_\times}{\partial e} \right\rvert\, \frac{\partial h_\times}{\partial e}\right\rangle_{\gamma_0} \nonumber\\ &\sim  2h^2 \cos ^2 \Theta  T_{\rm obs}  \sum_n \frac{1}{S_n(nf_{\text {orb }})} ( B_{n+}^{\prime 2} + B_{n-}^{\prime 2}) 
\end{align}
where a prime denotes the derivative, i.e., $B_{n\pm}^{\prime} \equiv dB_{n\pm}/de$. 

We apply the same analysis to the plus polarization (see Equation~(\ref{eq:eccwaveform})), and get:
\begin{align}
\left\langle\Gamma^{+}_{{f_{\rm orb}f_{\rm orb}}} \right\rangle_{\gamma_0} & =\left\langle\left.\frac{\partial h_+}{\partial f_{\text {orb }}} \right\rvert\, \frac{\partial h_+}{\partial f_{\text {orb }}}\right\rangle_{\gamma_0}  \nonumber\\ \sim \frac{2\pi^2}{3} h^2   T_{\rm obs}^3  &\sum_n \frac{n^2}{S_n(nf_{\text {orb }})} \\
&\times \left[\sin ^4 \Theta A_{n}^2+(1+\cos ^2 \Theta)^2( B_{n+}^2 + B_{n-}^2) \right]\nonumber\ ,
\end{align}
\begin{align}
\left\langle\Gamma^{+}_{\delta f \delta f} \right\rangle_{\gamma_0} &  =\left\langle\left.\frac{\partial h_+}{\partial \delta f} \right\rvert\, \frac{\partial h_+}{\partial \delta f}\right\rangle_{\gamma_0}  \\ &\sim \frac{2\pi^2}{3} h^2   T_{\rm obs}^3  \sum_n \frac{(1+\cos ^2 \Theta)^2}{S_n(nf_{\text {orb }})} ( B_{n+}^2 + B_{n-}^2)\nonumber \ . \label{eq:plusee}
\end{align}
\begin{align}
\left\langle\Gamma^{+}_{ee} \right\rangle_{\gamma_0} & =\left\langle\left.\frac{\partial h_+}{\partial e} \right\rvert\, \frac{\partial h_+}{\partial e}\right\rangle_{\gamma_0}  \nonumber\\ &\sim \frac{1}{2} h^2   T_{\rm obs}  \sum_n \frac{1}{S_n(nf_{\text {orb }})} \\
&\times \left[\sin ^4 \Theta A_{n}^{\prime 2}+(1+\cos ^2 \Theta)^2( B_{n+}^{\prime2} + B_{n-}^{\prime2}) \right]\nonumber
\end{align}

In realistic observation, the detector's output is a linear combination of two polarizations of the gravitational wave, and the coefficients depend on the sources' sky location and detector's orientation (see the discussions in Section~\ref{subsec:fishernum}. Here, for simplicity, we combine the matrix elements for $h_{\times},\,h_+ $ using:
\begin{equation}
    \Gamma=\Gamma^{\times}+\Gamma^{+}\label{eq:gammatotal}
\end{equation}

We note that, after taking the approximations in Equation~(\ref{eq:approximation}) and averaging the matrix elements over $\gamma_0$, the cross-terms between different parameters will vanish because of the same reason as discussed below Equations~(\ref{eq:selfproduct}) and (\ref{eq:crossterm}), i.e., 
\begin{equation}
    \left\langle\Gamma^{\times+}_{f_{\rm orb} \delta f}\right\rangle_{\gamma_0}\sim 0,\left\langle \Gamma^{\times+}_{f_{\rm orb} e}\right\rangle_{\gamma_0}\sim 0,\left\langle \Gamma^{\times+}_{\delta f e}\right\rangle_{\gamma_0} \sim 0\ ,\label{eq:nocrossterms}
\end{equation}
therefore, the diagonal terms in the inverse of the Fisher matrix, $C_{ii}$, can be simplified as $C_{i i} \sim \left\langle\Gamma_{ii} \right\rangle_{\gamma_0}^{-1}$, which gives the following relation (see Equation~(\ref{eq:delt_estimation})):
\begin{equation}
\Delta \lambda_{i}=C_{i i}^{\frac{1}{2}} \sim \left\langle\Gamma_{ii} \right\rangle_{\gamma_0}^{-\frac{1}{2}}\ .\label{eq:delta_result}
\end{equation}

We can plug Equations~(\ref{eq:gammaforbcorss}) - (\ref{eq:plusee}) into Equation~(\ref{eq:gammatotal}), then use Equation~(\ref{eq:delta_result}) to calculate the parameter measurement error:
\begin{align}\label{eq:dforb}
&\Delta f_{\rm orb}\sim \left(\left\langle \Gamma^{\times}_{f_{\rm orb} f_{\rm orb}}\right\rangle_{\gamma_0}+\left\langle \Gamma^{+}_{f_{\rm orb} f_{\rm orb}}\right\rangle_{\gamma_0} \right)^{-\frac{1}{2}}\nonumber\\
& = h^{-1}T_{\rm obs}^{-\frac{3}{2}}\left\{\frac{2\pi^2}{3} \sum_n \frac{n^2}{S_n(nf_{\text {orb }})} \right.\\
&{\left.\times \left[\sin ^4 \Theta A_{n}^2+(1+6\cos ^2 \Theta+\cos ^4 \Theta)( B_{n+}^2 + B_{n-}^2) \right] \right\}^{-\frac{1}{2}}}\nonumber\ ,
\end{align}
\begin{align}\label{eq:ddf}
& \Delta (\delta f) \sim \left(\left\langle \Gamma^{\times}_{\delta f \delta f}\right\rangle_{\gamma_0}+\left\langle \Gamma^{+}_{\delta f \delta f}\right\rangle_{\gamma_0} \right)^{-\frac{1}{2}} \\
& = h^{-1}T_{\rm obs}^{-\frac{3}{2}}\left\{\frac{2\pi^2}{3} \sum_n \frac{(1+6\cos ^2 \Theta+\cos ^4 \Theta)}{S_n(nf_{\text {orb }})} ( B_{n+}^2 + B_{n-}^2) \right\}^{-\frac{1}{2}}\nonumber\ ,
\end{align}
\begin{align}\label{eq:de}
&\Delta e\sim \left(\left\langle \Gamma^{\times}_{ee}\right\rangle_{\gamma_0}+\left\langle \Gamma^{+}_{ee}\right\rangle_{\gamma_0} \right)^{-\frac{1}{2}}\nonumber\\
& = h^{-1}T_{\rm obs}^{-\frac{1}{2}}\left\{ \frac{1}{2}\sum_n \frac{1}{S_n(nf_{\text {orb }})} \right.\\
&{\left.\times \left[\sin ^4 \Theta A_{n}^{\prime2}+(1+6\cos ^2 \Theta+\cos ^4 \Theta)( B_{n+}^{\prime2} + B_{n-}^{\prime2}) \right] \right\}^{-\frac{1}{2}}}\nonumber\ .
\end{align}

We emphasize that, Equations~(\ref{eq:dforb}) - (\ref{eq:de}) serve as a lower bound of the measurement error since we exclude $(h,\Theta,\gamma_0)$ when analyzing the elements of the Fisher matrix. In fact, it turns out that this estimation may differ from the numerical results by orders of magnitude (see, e.g., Figure~\ref{fig:fishernum}). However, we expect it to show the correct dependence of $\Delta f_{\rm orb}$, $\Delta (\delta f)$, and $\Delta e$ on the observational time, as well as the orbital parameters of the binary.

\section{\xzy{Computing the Major and Minor Axes of Sky Error Ellipsoids}}
\label{app:skyerror}

\xzy{To estimate the parameter measurement errors for GW sources, we computed their Fisher matrix using the partial derivatives of the waveform with respect to 10 parameters: $f_{\rm orb}, 1-e_0, M, q, \cos\Theta, \Phi, \cos\theta,\phi, R,\psi$ (see Section~\ref{subsec:fishernum}). On the other hand, instead of directly estimating the errors in the spherical polar angles $\{\cos\Theta, \Phi, \cos\theta, \phi\}$, the source orientation and sky location measurement accuracy can be better described using sky error ellipsoids \citep[see, e.g.,][]{Lang2006,Kocsis_2008,Kocsis2007,Mikoczi+12}. Therefore, this section will briefly summarize how to convert the Fisher matrix analysis results of $\{\cos\Theta, \Phi, \cos\theta,\phi\}$ into the major and minor axes of the corresponding sky error ellipsoids, $\{2A_s, 2B_s\}$,$\{ 2a_s, 2b_s\}$.}

\xzy{In particular, we follow the method described in Sec. VI C of  Ref.~\citep{Lang2006}. For example, to estimate the error ellipsoid of sky location, $\{2a_s, 2b_s\}$, we first consider the relevant terms in the covariance matrix, $C=F^{-1}$, (see Equations~(\ref{eq:fisherdefinition}) and (\ref{eq:delt_estimation})):
\begin{equation}
\left\{
\begin{aligned}
&\Sigma_{\cos\theta,\cos\theta}  =C_{77}, \\
&\Sigma_{\cos\theta,\phi}  =\Sigma_{\phi,\cos\theta}=C_{78}, \\
&\Sigma_{\phi,\phi}  =C_{88},
\end{aligned}
\label{eq:errelps1}
\right.
\end{equation}
where indices 7 and 8 correspond to the parameters ${\cos\theta, \phi}$ in the Fisher matrix $F_{ij}$ and covariance matrix $C_{ij}$.}

\xzy{Defining the error ellipse such that the probability of the source lying outside the ellipse is $e^{-1}$ \citep{Cutler+98}, the major axis $2 a_s$ and minor axis $2 b_s$ of the ellipse are given by
\begin{equation}
\begin{aligned}
& 2\left[\csc ^2 \theta \Sigma_{\cos\theta,\cos\theta}+\sin ^2 \theta \Sigma_{\phi,\phi} \pm\right. \\
& \left.\sqrt{\left(\csc ^2 \theta \Sigma_{\cos\theta,\cos\theta}-\sin ^2 \theta \Sigma_{\phi,\phi}\right)^2+4\left(\Sigma_{\cos\theta,\phi}\right)^2}\right]^{1 / 2}
\end{aligned}
\label{eq:errelps2}
\end{equation}}

\xzy{Similarly, this method can be applied to the parameters $\{\cos\Theta, \Phi\}$, by substituting $\cos\theta\rightarrow\cos\Theta$, $\phi\rightarrow\Phi$ in Equations~(\ref{eq:errelps1}) and (\ref{eq:errelps2}), yielding the major and minor axes of error ellipse for source orientation, $\{2A_s, 2B_s\}$.}


\begin{figure*}[htbp]
    \centering
    \includegraphics[width=7.5in]{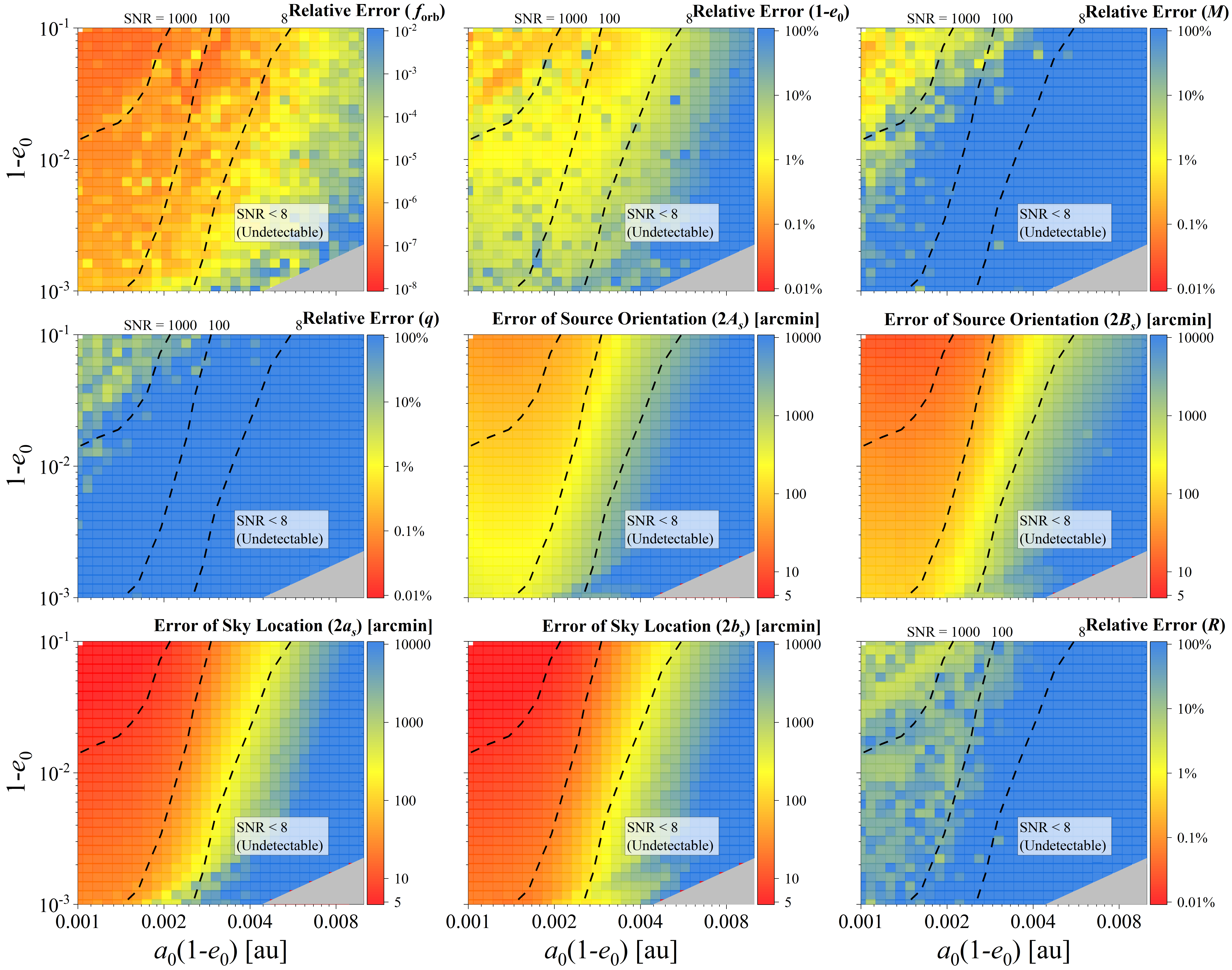} 
    \caption{{\bf{The dependence of a compact binary's parameter measurement error on its initial pericenter distance, $r_p= a_0(1-e_0)$, and eccentricity $e_0$. (1-yr observation) }}
    Here we consider the same systems as in Figure~\ref{fig:fishernum}, but re-parameterize the x-axis using $r_p= a_0(1-e_0)$. We exclude the grey-colored region at the bottom of each panel since here the binary's orbital period is longer than the observation time (1 year).
    }
    \label{fig:fishernum2}
\end{figure*}
\begin{figure*}[t]
    \centering
    \includegraphics[width=3.5in]{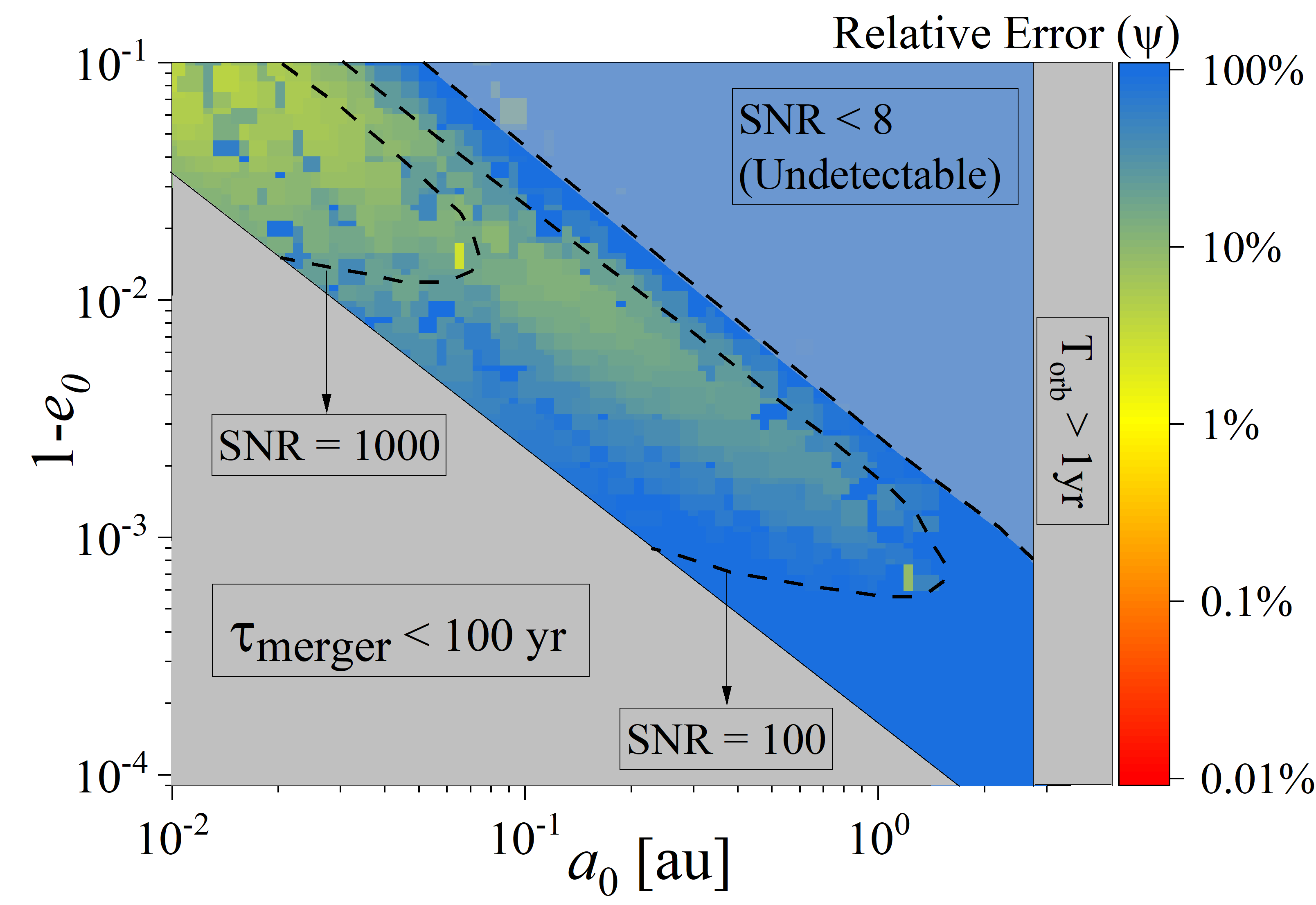} 
    \caption{{\bf{The dependence of a compact binary's polarization angle measurement error on its semi-major axis and eccentricity, computed using the Fisher matrix analysis. (1-yr observation) }} Here we consider the same systems as in Figure~\ref{fig:fishernum}, but show the result according to parameter $\psi$ for completeness.
    }
    \label{fig:gcbkgeg}
\end{figure*}


\begin{figure*}[t]
    \centering
    \includegraphics[width=3.5in]{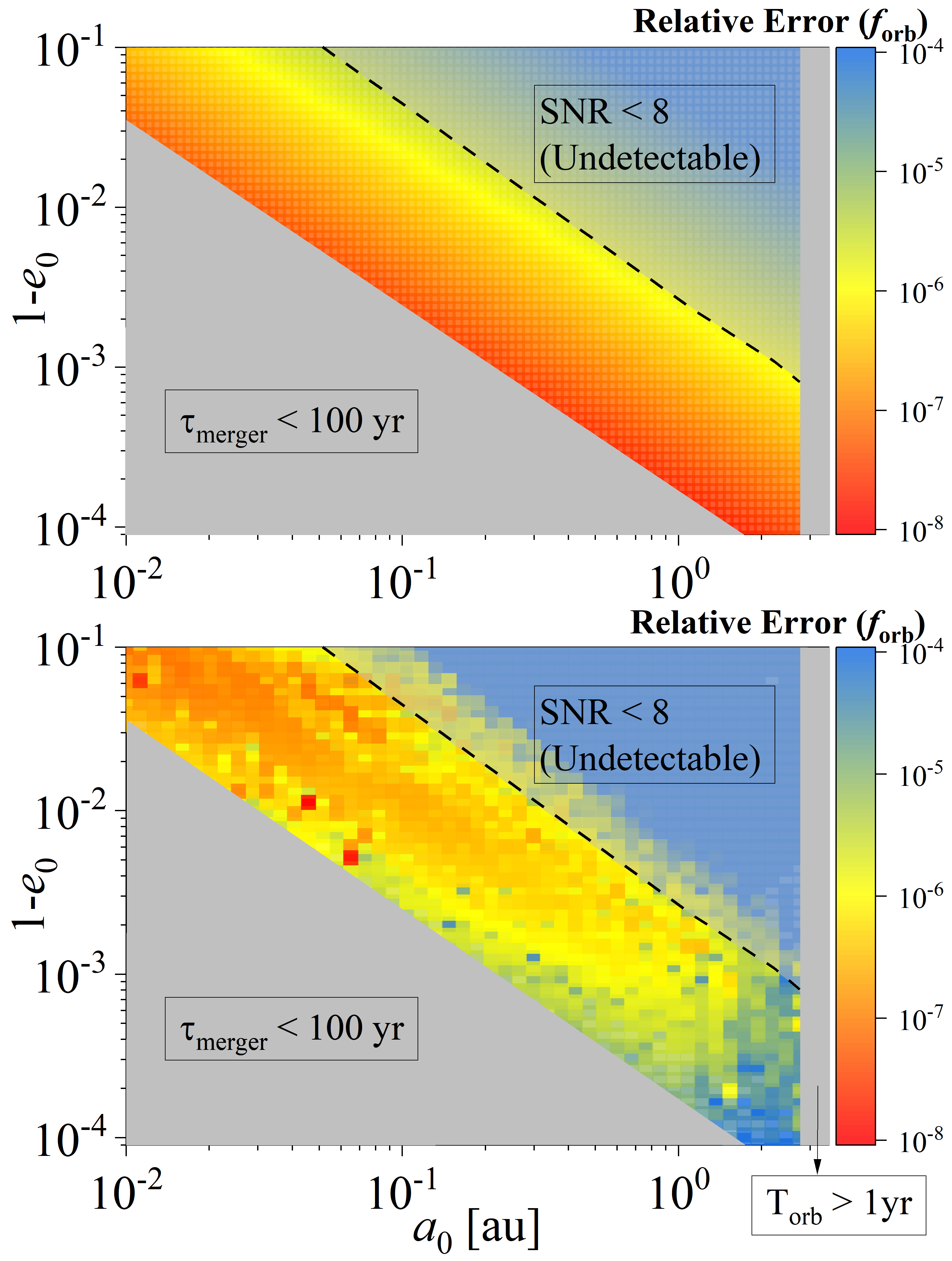} 
    \caption{{\bf Comparison between the analytical and numerical estimation of $f_{\rm orb}$ measurement error.} Here we consider the same systems as shown in Figure~\ref{fig:fishernum}, but calculate the relative $f_{\rm orb}$ measurement error using Equation~(\ref{eq:analyticalestiforb}) ({\it Upper Panel} ), and compare it with the numerical results from Fisher matrix analysis ({\it Bottom Panel} ). As shown in the figure, the analytical approach provides an accurate estimation of $f_{\rm orb}$ measurement error in the moderate $\rm SNR$ region (i.e., close to the dashed line of $\rm SNR=8$). However, its performance is limited when the binary has a short merger timescale (to the bottom left of each panel) and becomes inconsistent with the Fisher matrix analysis.
    }
    \label{fig:fisherana}
\end{figure*}

\end{document}